\documentclass[sigconf]{acmart}
\AtBeginDocument{%
  }

\setcopyright{acmlicensed}
\copyrightyear{2026}
\acmYear{2026}
\acmDOI{XXXXXXX.XXXXXXX}
\acmConference[Conference acronym 'XX]{Make sure to enter the correct
  conference title from your rights confirmation email}{June 03--05,
  2018}{Woodstock, NY}
\acmISBN{978-1-4503-XXXX-X/2026/06}

\usepackage{multirow}
\usepackage{todonotes}
\usepackage{graphicx}
\usepackage{float}
\usepackage{subfigure}
\usepackage{makecell}
\usepackage[most]{tcolorbox}
\usepackage{threeparttable}
\usepackage{hyphenat}
\usepackage{float}
\usepackage{placeins}
\usepackage[normalem]{ulem}

\usepackage{xparse}

\begin{document}

\title{Hieronym: Leveraging Hierarchical Multi-Source Information for Function Renaming in Stripped Binary}

\author{Xiaoling Zhang}
\affiliation{%
  \institution{Zhongguancun Laboratory}
  \city{Beijing}
  \country{China}}
\email{zhangxl@zgclab.edu.cn}

\author{Jian Sun}
\affiliation{%
  \institution{Zhongguancun Laboratory}
  \city{Beijing}
  \country{China}}
\email{sunjian@zgclab.edu.cn}

\author{Dawei Wang}
\authornote{Corresponding authors.}
\affiliation{%
  \institution{Zhongguancun Laboratory}
  \city{Beijing}
  \country{China}}
\email{wangdw@zgclab.edu.cn}

\author{Chongyu Wang}
\affiliation{%
  \institution{Zhongguancun Laboratory}
  \city{Beijing}
  \country{China}}
\email{wangchy@zgclab.edu.cn}

\author{Li Chen}
\authornotemark[1]
\affiliation{%
  \institution{Zhongguancun Laboratory}
  \city{Beijing}
  \country{China}}
\email{lichen@zgclab.edu.cn}

\author{Zhaoteng Yan}
\affiliation{%
  \institution{Zhongguancun Laboratory}
  \city{Beijing}
  \country{China}}
\email{yanzt@zgclab.edu.cn}

\author{Peipei Liu}
\affiliation{%
  \institution{Zhongguancun Laboratory}
  \city{Beijing}
  \country{China}}
\email{liupp@zgclab.edu.cn}

\author{Lixiao Zhang}
\affiliation{%
  \institution{Zhongguancun Laboratory}
  \city{Beijing}
  \country{China}
  }
\affiliation{ \institution{Harbin Institute of Technology}
  \city{Harbin}
  \country{China}
  }
\email{zhanglx2025@zgclab.edu.cn}

\author{Dan Li}
\affiliation{%
  \institution{Tsinghua University}
  \city{Beijing}
  \country{China}}
\email{tolidan@tsinghua.edu.cn}

\renewcommand{\shortauthors}{Xiaoling Zhang et al.}

\begin{abstract}
Function renaming in stripped binaries can substantially assist reverse engineers by improving code readability, yet it is a challenging task. The difficulty stems from the need to accurately capture function semantics from low-level binary code across diverse instruction sets, architectures, and compiler optimizations, and to express these semantics in concise, human-readable names. Existing approaches either inadequately capture comprehensive function semantics or exhibit limited generalization to previously unseen binaries.
In this paper, we present \textsc{Hieronym}, a generative large language model (LLM)-based framework for stripped binary function renaming. \textsc{Hieronym} adopts a hierarchical summarization-driven domain adaptation strategy and integrates multi-source information—including global binary context, local calling context, and intrinsic function semantics—to enhance the LLM’s understanding of binary code. To enable systematic evaluation, we further propose a dual-layer evaluation framework that incorporates both token-level and whole-name-level metrics. 
We evaluate \textsc{Hieronym} on binary functions compiled with four compiler optimization levels (O0–O3) for four architectures (x64, x86, ARM, and MIPS).  
Experimental results demonstrate that \textsc{Hieronym} significantly outperforms state-of-the-art methods, achieving token-level improvements of 50.12\% in precision, 41.75\% in recall, and 45.10\% in F1-score, as well as a 79.94\% improvement in name-level accuracy, while also exhibiting strong generalization capability.
Moreover, experiments on real-world malware samples further validate the practical effectiveness of \textsc{Hieronym} in security-critical scenarios.

\end{abstract}


\maketitle
\section{Introduction}
Reverse engineering is a fundamental technique in security analysis, supporting tasks such as vulnerability discovery~\cite{10179479,10.1145/3597926.3598121}, malware analysis~\cite{285445,10.1145/3176258.3176335}, and security auditing~\cite{235499}.
In practice, analysts frequently encounter stripped binaries lacking high-level information, including function and variable names, structure details and optimization information. This loss obscures program intent and hinders analysis efficiency.
Among the missing information, function names are particularly critical, as they provide concise abstractions of program functionality and serve as key cognitive anchors during program comprehension~\cite{Votipka,taxonomy}. 
However, widely used decompilation tools such as IDA Pro and Ghidra typically assign address-based placeholder names (such as \texttt{FUN\_00129f48}) that convey little information. Accurate function naming still depends heavily on manual expert analysis~\cite{ZHANG2025112492}, rendering the process time-consuming and labor-intensive.
Consequently, automatic function renaming for stripped binaries has emerged as an important research direction.

However, automatic function renaming is a challenging task, as it requires both a comprehensive understanding of binary function semantics and effective abstraction of such semantics into concise and meaningful natural-language keywords.
Recent machine learning (ML)-based approaches have shown promise~\cite{Debin, NFRE, Prob_name, symlm, patrick2023xfl, Epitome}. 
These methods typically employ weak base models, such as LSTM~\cite{10.1007/s10772-018-09573-7} in NFRE~\cite{NFRE} and BERT~\cite{b38} in SymLM~\cite{symlm}, to understand function semantics and adopt a classification-based paradigm that maps function semantics to a predefined, closed vocabulary.
These models often struggle to capture function semantics robustly, and exhibit limited ability to abstract function semantics into natural‑language keywords. While they may perform well on training data, they often lack sufficient generalization to unseen binaries. 
Generative large language models (LLMs) have recently demonstrated strong generalization and reasoning capabilities, achieving impressive results in code understanding and generation. These advances have spurred growing interest in applying LLMs to binary function renaming. However, despite achieving state-of-the-art (SOTA) performance, existing LLM-based methods still fall short of practical expectations: even the best-performing approach achieves only a 0.277 F1-score. This discrepancy raises a critical question: \textbf{why have existing methods not delivered the performance we would expect from LLMs in this setting?} We analyze this gap and identify the following challenges.

\noindent\textbf{C1: Flawed Evaluation Metrics.} 
Prior evaluation metrics for function renaming are flawed, as they sometimes require models to reproduce fine-grained details unavailable from binary function alone.
For example, the prefix in \texttt{jitterc\_get\_lineno} encodes project-specific context unrecoverable from the binary functions alone. 
Penalizing the prediction of \texttt{get\_current\_line\_number} is therefore unjustified.
Moreover, most evaluations rely on token-level metrics to mitigate the out-of-vocabulary (OOV) issue, but this choice is fragile. Erroneous tokenization can drastically distort evaluation metrics. For instance, SymGen~\cite{symgen} tokenizes \texttt{c\_isascii\_string} into \{\texttt{c}, \texttt{isa}, \texttt{sc}, \texttt{ii}, \texttt{string}\}, reducing its F1-score from 0.857 to 0.25. 
Since a sound evaluation is essential for revealing the true limitations of existing methods and supporting a faithful interpretation of their performance, how to properly evaluate the performance of the function renaming task is a non-trivial challenge.

\noindent\textbf{C2: Semantic Sparsity of Stripped Binaries.}
Motivated by \textbf{C1}, we observe that stripped binary functions contain only limited semantic information, which often makes accurate name recovery unattainable. For example, the \texttt{jitterc}-specific identifier discussed above is essentially impossible to recover from a single stripped binary function. Although prior works, such as SymLM~\cite{symlm}, XFL~\cite{patrick2023xfl}, and BLens~\cite{blen}, attempt to incorporate additional context, such as calling context, this information is often still insufficient to compensate for the loss of semantics in stripped binaries, especially for context-specific identifiers. 
Moreover, the contextual representations used by these methods are sensitive to architectural and optimization variations, which limits their robustness. Therefore, a key challenge is how to model essential, task-relevant context from information-limited binaries while remaining robust to architecture- and optimization-induced changes.

\noindent\textbf{C3: Knowledge Gaps Between Natural Language and Binary Code Semantics in LLMs.}
Although LLMs are strong at understanding and generating source code, their training corpora are dominated by natural language and include relatively little code, which leaves them with limited ability to understand stripped binary~\cite{touvron2023llamaopenefficientfoundation,Radford2019LanguageMA}.
This mismatch creates a substantial gap between the knowledge encoded in general-purpose LLMs and the understanding required for binary analysis.
A straightforward approach to mitigating this gap is domain adaptation. Accordingly, function summarization-based domain adaptation strategies have been explored and shown some effectiveness in transferring LLMs to the binary domain~\cite{symgen}. However, prior studies indicate that function-level summaries alone are insufficient to capture deep and comprehensive function semantics~\cite{pascarella2019classifying,HUANG2020106373,huang2019learning,10.1145/3611664,10.1145/3582570}. Consequently, effectively bridging the knowledge gap between general-purpose LLMs and the binary domain remains an urgent open problem.

In this paper, we present \textsc{Hieronym}, a novel generative framework for function renaming in stripped binaries. 
It is designed to learn hierarchical binary function semantics and to comprehensively model binary functions by integrating multi-source information. Leveraging generative LLMs, \textsc{Hieronym} abstracts binary semantics into natural-language keywords.
To ensure comprehensive and reliable evaluation (\textbf{C1}), we adopt a dual-layer evaluation framework that assesses performance at both the token and name levels. We propose an enhanced token-level evaluation with a hybrid tokenization mechanism to produce semantically meaningful tokens to mitigate tokenization errors, along with heuristic rules to measure token-level semantic relatedness. Our evaluation framework is further completed by a name-level evaluation using an LLM-as-a-judge approach to assess whole-name semantic equivalence.
To comprehensively model function semantics (\textbf{C2}), \textsc{Hieronym} integrates three complementary information sources: global binary context, local calling context, and the target function itself.
Global and local contexts provide essential task-relevant information, which we represent using textual summaries. 
Our representation offers a robust abstraction that is resilient to code variations introduced by different compilation.
To bridge the knowledge gap between general-purpose LLMs and the binary domain (\textbf{C3}), we introduce hierarchical function summarization as a domain adaptation strategy, enabling multi-view semantic understanding of functions.

Our comprehensive evaluation demonstrates that \textsc{Hieronym} exhibits exceptional generalizability and superior performance compared to SOTA function renaming methods, including SymLM~\cite{symlm}, XFL~\cite{patrick2023xfl}, BLens~\cite{blen}, and SymGen~\cite{symgen}, across four instruction set architectures (x64, x86, ARM, and MIPS) and four optimization levels (O0–O3). 
\textsc{Hieronym} achieves substantial gains in token-level precision, recall, and F1-score of 50.12\%, 41.75\%, and 45.10\%, respectively, as well as a 79.94\% improvement in name-level accuracy over the best prior method. Leveraging the multi-source information in our approach, \textsc{Hieronym} even successfully recovers the \texttt{jitterc}-specific identifier discussed above.
Ablation studies further validate the effectiveness of \textsc{Hieronym}’s design. Hierarchical function summarization based domain adaptation improves F1-score by 18.83\%, while incorporating global binary context and local calling context yields additional gains of 27.63\% and 13.81\%, respectively. These results highlight the contributions of each component and underscore the novelty and effectiveness of the proposed framework.
Finally, experiments on real-world malware binaries confirm \textsc{Hieronym}’s practical utility for security analysis.

The contributions of this paper can be summarized as follows:

$\bullet$ We propose a dual-layer evaluation framework that evaluates performance at both the token-level and the name-level, enabling a reliable evaluation of function renaming methods.

$\bullet$ We propose \textsc{Hieronym}, a novel generative approach for function renaming in stripped binaries that leverages multi-source information. By integrating global binary context, local calling context, and target function, \textsc{Hieronym} enables LLMs to learn comprehensive function semantics, resulting in improved performance and generalizability.
    
$\bullet$ We introduce a hierarchical function summarization-based domain adaptation strategy that effectively adapts pre-trained generative LLMs to binary semantic modeling.

$\bullet$ Extensive experiments show that \textsc{Hieronym} consistently outperforms SOTA methods in function renaming. The results further highlight its strong generalizability, the effectiveness of individual components, and its practical applicability. We make our code and dataset publicly available at: \url{https://github.com/NASP-THU/Hieronym}.

\section{Background and Motivation}
\subsection{Problem Definition}
Function renaming in stripped binaries requires both accurate reverse engineering of binary semantics and the generation of names that faithfully reflect those semantics. Decoder-only LLMs, based on the Transformer decoder architecture~\cite{Radford2019LanguageMA,touvron2023llamaopenefficientfoundation}, have demonstrated strong capabilities in text understanding and generation, particularly exhibiting impressive generalization in zero- and few-shot settings~\cite{10.5555/3495724.3495883}.
In contrast to existing ML-based approaches~\cite{NFRE,symlm,Epitome}, which formulate function renaming as a classification task over a closed vocabulary, we follow SymGen~\cite{symgen} and model function renaming as an autoregressive generation task using LLMs. 
Formally, given a binary function $\mathcal{F}$, represented as a sequence of binary code tokens $\mathcal{F} = \{t_1, t_2, \ldots, t_n\}$, a function renaming model $\mathcal{M}$ generates a sequence of natural-language tokens $\mathcal{W} = \{w_1, w_2, \ldots, w_m\}$, i.e., $\mathcal{M}: \mathcal{F} \rightarrow \mathcal{W}$. Under the autoregressive paradigm, the generation probability of $\mathcal{W}$ is factorized as
\begin{equation}
p(\mathcal{W}) = \prod_{j=1}^{m} p(w_j \mid w_1, \ldots, w_{j-1}, \mathcal{F}),
\end{equation}
where each token $w_j$ is conditioned on all previously generated tokens and the binary function semantics.

\begin{figure}[htbp]
    \centering
    \includegraphics[width=\linewidth]{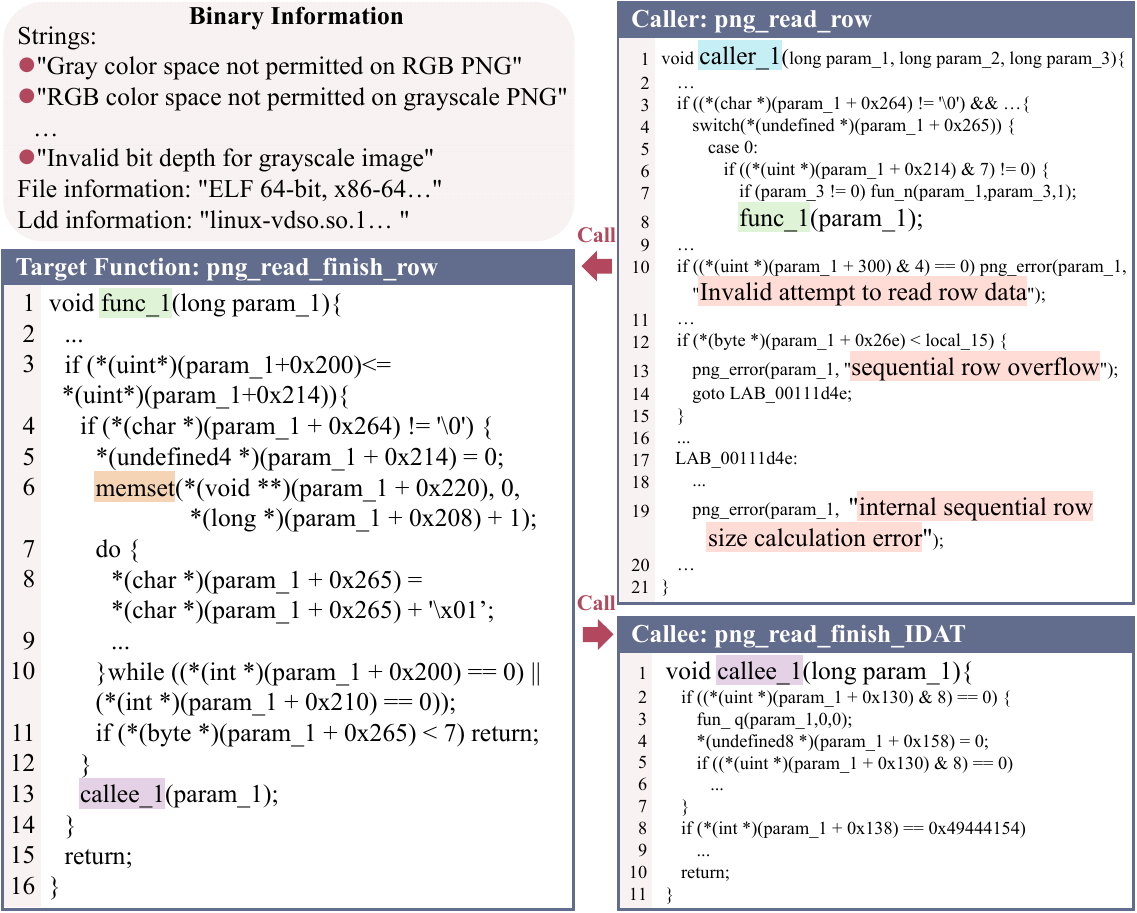}
    \caption{An Example Illustrating the Importance of Task-Related Information in Function Naming}
    
    \label{fig:motivation}
\end{figure}

\subsection{Existing Techniques and Our Motivations} \label{motivation}
Although many studies have explored function renaming in stripped binaries, prior efforts have not adequately addressed the challenges outlined above, which directly motivates our proposal of \textsc{Hieronym}.

Existing methods primarily evaluate function-renaming performance at the token level, yet the reliability of such evaluation heavily depends on the underlying tokenization strategy, which directly affects subsequent semantic alignment.
Many methods—including XFL~\cite{patrick2023xfl}, BLens~\cite{blen}, SymLM~\cite{symlm}, and SymGen~\cite{symgen}—rely on rule-based tokenization, which often introduces segmentation bias and degrades evaluation metrics. For example, SymLM incorrectly splits \texttt{test\_nofork\_sideffects} into \{\texttt{t}, \texttt{est}, \texttt{no}, \texttt{for}, \texttt{k}, \texttt{side}, \texttt{eff}, \texttt{ect}, \texttt{s}\}, hindering correct alignment with the ground-truth tokens.
Semantic alignment based on tokenization results also remains limited. 
XFL and BLens depend on general English synonym sets that fail to capture domain-specific morphological variations, while the CodeWordNet~\cite{symlm} models used by SymLM and SymGen are trained on small corpora with insufficient coverage of OOV tokens.
As a result, existing token-level evaluation strategies suffer from unreliable tokenization and weak semantic alignment, and the absence of name-level evaluation further prevents current methods from systematically addressing \textbf{C1}.

Grasping task-relevant contextual information is essential for modeling function semantics and improving function-renaming performance. Figure~\ref{fig:motivation} illustrates this through a real-world example of a stripped binary, showing the decompiled code of a target function together with its callers, callees, and domain-related information extracted from binary (simplified for clarity).
The example involves three functions—\texttt{func\_1}, \texttt{caller\_1}, and \texttt{callee\_1}—whose ground-truth names are \texttt{png\_read\_finish\_row}, \texttt{png\_read\_row}, and \texttt{png\_read\_finish\_IDAT}, respectively. Although decompiled code exposes richer semantics than raw binaries, the target function alone is insufficient for accurate name inference.
As shown in Figure~\ref{fig:motivation}, the semantics of \texttt{png\_read\_finish\_row} are tightly coupled with both its caller and callee. 
Prior works have attempted to incorporate calling context in different ways. 
SymLM models calling context using assembly-level representations, while XFL and BLens rely on handcrafted features to capture calling relationships. 
However, these methods are sensitive to architecture and optimization settings, causing unstable semantics (e.g., SymLM reports a 60.2\% F1-score drop from O3 to O0).
Domain-related information (e.g., the keyword “PNG”) provides essential task-relevant cues unavailable from the local function alone.
Despite its importance, existing methods do not consider such domain-related information. 
The LLM-based method SymGen ignores both domain information and calling context. 
Existing methods lack effective mechanisms for leveraging task-relevant information and struggle to address \textbf{C2}.

Many existing methods, such as SymLM, BLens, and XFL, are built on traditional ML frameworks and lack intrinsic mechanisms to address \textbf{C3}. 
SymGen introduces function-level summarization to equip LLMs with binary code knowledge. However, prior studies show that function-level summaries alone are insufficient to capture fine-grained implementation details. 
In particular, statement-level summarization has been shown to effectively complement function-level semantics by modeling the distinct logical operations within a function~\cite{HUANG2020106373,10.1145/3611664}.
These two levels of abstraction are inherently complementary: statement-level summaries capture concrete implementation logic, while function-level summaries convey overall functionality. Together, they provide a more complete and nuanced understanding of code structure and semantics~\cite{10.1145/3611664,10.1145/3582570}. As a result, SymGen remains limited in modeling fine-grained binary semantics and does not fully address \textbf{C3}.
\section{Evaluation Framework}
To comprehensively evaluate function renaming methods, we propose a dual-level evaluation framework: (1) \textbf{token-level evaluation}, which is widely used in prior work to mitigate OOV issue;
and (2) \textbf{name-level evaluation}, which measures the correctness of generated function names from a holistic semantic perspective. 
By combining these two complementary levels, our framework enables a systematic evaluation of both fine-grained semantic recognition and overall semantic expressiveness.

\subsection{Token-Level Evaluation} \label{sec:evaluation}
Token-level evaluation typically consists of two stages: (1) function-name tokenization, which segments a name into meaningful tokens, and (2) token semantic alignment, which determines semantic relationships among tokens. In the tokenization stage, we adopt a hybrid strategy that combines rule-based heuristics with LLM assistance. In the subsequent alignment stage, heuristic rules are applied to identify semantic associations between tokens.

\noindent \textbf{Name Tokenization.}
Function names are typically multi-word natural-language expressions that include abbreviations, domain-specific jargon, and language-dependent conventions~\cite{hindle2016naturalness,scanniello2017fixing}. 
Unlike natural language text with explicit word boundaries, code supports diverse naming styles. 
Although capitalization, underscores, or digits are commonly used as delimiters, they are often omitted, resulting in unsegmented character sequences with ambiguous word boundaries.
For function names that follow conventional naming styles—such as camelCase, underscore-separated, or numeric-delimited forms—we apply a rule-based tokenization strategy. For unconventional names, rule-based methods used in prior work~\cite{symlm,symgen} tend to over-segment names, breaking coherent semantic tokens and thereby undermining token-level evaluation.
To address this limitation, we introduce an LLM-based few-shot tokenization approach that leverages contextual understanding to achieve accurate segmentation of complex function names. The prompt is illustrated in Figure~\ref{fig:prompt5} in Appendix~\ref{tokenization}. We provide the LLM with a small set of diverse in-context examples covering different naming styles and domain-related vocabulary to guide structured parsing.
For each function name, we generate three independent tokenization results and aggregate them using a voting-based strategy. The final segmentation is determined as follows:

$\bullet$ A segmentation boundary is retained if it is suggested in at least two of the three results.

$\bullet$ Among the three candidates, the result with the largest intersection with the retained boundary set is selected, with any redundant or overlapping labels removed.
 
By integrating rule-based heuristics with LLM-driven tokenization, our framework achieves robust and effective segmentation of function names, providing a reliable foundation for subsequent token-level semantic analysis. 

\noindent \textbf{Token Semantic Alignment.}
Function names exhibit substantial lexical variability due to developer preferences, including morphological variants, abbreviations, and misspellings~\cite{jiang2019semantic,hindle2016naturalness,scanniello2017fixing,symlm}. Ignoring such relationships significantly biases evaluation, and manual curation does not scale to large vocabularies.
To address this issue, we build on prior work~\cite{NFRE,Epitome} and propose an automated method for identifying semantically related tokens. We categorize token relationships into four types and handle each as follows:

$\bullet$ For morphological variants, we apply lemmatization to determine whether two tokens share the same lexical root.

$\bullet$ For abbreviation, we use Smith-Waterman algorithm~\cite{smith1981identification} to compute token similarity. If the relative similarity exceeds 0.65 and both tokens are longer than two characters, they are considered to form an abbreviation-full-form relationship.

$\bullet$ For misspellings, we propose a Damerau-Levenshtein-based measure~\cite{10.5555/1274531.1274545} to quantify token similarity, pairs exceeding a threshold of 0.75 are classified as spelling variants.

$\bullet$ For synonyms, we use Qwen3-Embedding~\cite{qwen3embedding} to mitigate OOV issues, trained on large-scale corpora, pairs with similarity scores above 0.8 are treated as synonyms.

By incorporating multi-dimensional association mechanism, our evaluation framework can identify semantically equivalent tokens, improving evaluation robustness in the presence of lexical diversity. The thresholds used in this mechanism are determined through empirical analysis and validated via experimental evaluation.

\begin{figure*}[tb]
    \centering
    \vspace{-8pt}
    \includegraphics[scale=0.5]{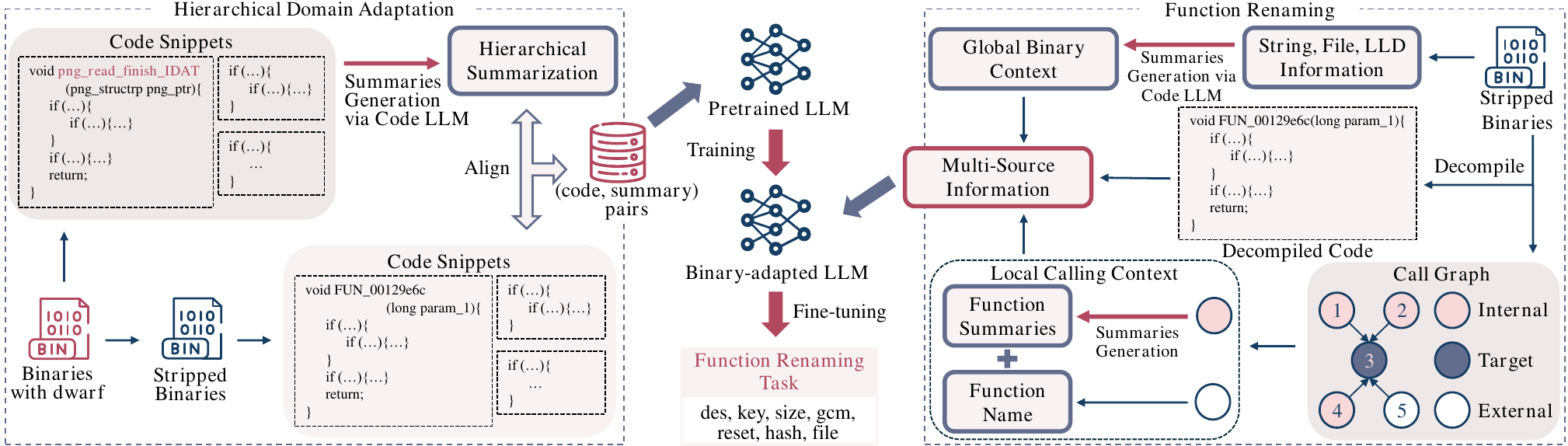}
    \caption{The Overall Workflow of \textsc{Hieronym}}
    \Description{The Overall Workflow of \textsc{Hieronym} }
    \label{fig:arch}
\end{figure*}

\noindent \textbf{Token-Level Metrics.}
To evaluate our model and existing methods, we adopt the standard token-level precision, recall, and F1-score metrics, following common practice in prior studies~\cite{symlm,Epitome,symgen,blen}. These metrics quantify the overlap between tokens in the predicted and ground-truth function names.
We define \emph{true positives} (\textit{TP}) as the number of correctly predicted tokens, including both exact matches and those considered equivalent under semantic alignment. \emph{False positives} (\textit{FP}) are the number of predicted tokens that do not appear in ground-truth tokens, while \emph{false negatives} (\textit{FN}) are the number of ground-truth tokens that are not predicted. 
Formally, given a predicted name $ \hat{W}: \{\hat{w_1}, \ldots, \hat{w_j}\}$ and the ground truth $W: \{w_1, \ldots, w_m\}$, where $w_t (\hat{w_t})$ denotes a token in the name, and $\alpha$ is an indicator function such that $\alpha \{True\} =1$ and $\alpha \{False\} = 0$, we compute:
\begin{equation*}
    TP = \sum_{t=1}^j \alpha {(\hat{w_t} \in W)} \quad
    FP = \sum_{t=1}^j \alpha {(\hat{w_t} \not\in W)} \quad
    FN = \sum_{t=1}^m \alpha {({w_t} \not\in \hat{W})}
    \label{TP}
\end{equation*}
Precision (P), recall (R), and F1-score (F1) are then computed as:
\begin{equation*}
    P = \frac{TP}{TP+FP} \qquad 
    R = \frac{TP}{TP+FN} \qquad
    F1 = \frac{2 \times P \times R}{P + R}
\end{equation*}

\subsection{Name–Level Evaluation}
Relying solely on token-level evaluation is insufficient for evaluating semantic consistency between predicted and ground-truth function names. 
Prior studies show that even for identical program semantics, developers choose the same identifier with a probability of only 6.9\%~\cite{feitelson2020developers}.
Function names may be lexically dissimilar yet semantically equivalent, causing token-level metrics (e.g., F1-score) to underestimate model performance.
we introduce a name-level evaluation that assesses semantic consistency from a holistic perspective to address this limitation. 
Recent studies indicate that LLMs used as evaluators align more closely with human judgment~\cite{su2024sourcecodefoundationmodels,xu2025unleashing}. Inspired by this finding, we adopt an LLM-assisted evaluation strategy and employ GPT-4o~\cite{openai_gpt4o_2024} as a semantic judge. 
For each binary function, we query GPT-4o with the global binary information, its decompiled code, and a pair consisting of the ground-truth name and the predicted name. 
GPT-4o is instructed to determine whether the predicted name is semantically consistent with the ground-truth name and to output a binary decision (“yes” or “no”).  The detailed prompt is provided in Figure~\ref{fig:prompt8} of Appendix~\ref{tokenization}.

To mitigate potential hallucinations or subjectivity in LLM-based evaluation, we further incorporate an embedding-based semantic similarity check. We compute the cosine similarity between the predicted and ground-truth names using the Qwen3-Embedding model and apply a threshold of 0.7 determined through empirical validation. 
A prediction is considered semantically correct only if both the GPT-4o judgment and the embedding-based similarity criterion indicate agreement.

\noindent \textbf{Name-Level Metric.}
We evaluate name-level performance using accuracy, defined as the proportion of predicted function names judged correct under the above criteria. Due to the scale of the dataset and the cost of LLM-based evaluation, we follow common practice~\cite{xu2025unleashing,Jin2023BinaryCS} and randomly sample 200 test functions for each architecture–optimization pair, yielding a total of 3,200 functions for name-level evaluation. This sampling strategy balances statistical reliability with computational feasibility.

\section{Approach}

\subsection{Overview}
Reverse-engineering practice indicates that analysts typically develop a global overview of a binary before examining functions in detail~\cite{Votipka, mantovani2022re}, such as its domain or functional scope. Motivated by this coarse-to-fine reasoning process and by the challenges discussed above, we propose \textsc{Hieronym}, a function renaming framework that leverages hierarchical code comprehension and multi-source information to robustly handle binaries compiled under diverse optimization levels. \textsc{Hieronym} consists of three key steps:
(I) \textbf{Hierarchical Domain Adaptation}, which enhances LLMs’ understanding of binary semantics at multiple granularities;
(II) \textbf{Multi-Source Information Fusion}, which integrates global binary context, local calling context, and the target function to construct a comprehensive semantic representation; and
(III) \textbf{Function Renaming Fine-Tuning}, which applies the fused information to the domain-adapted model to enable accurate and generalizable function renaming.

\subsection{Hierarchical Domain Adaptation} \label{statement}
We build a binary function renaming model upon a pre-trained generative LLM, leveraging its strong semantic modeling capability to capture complex binary function semantics and generalize across diverse binaries. Although general-purpose LLMs are exposed to a limited amount of source code during pre-training (e.g., approximately 5\% in the LLaMA dataset~\cite{touvron2023llamaopenefficientfoundation}), a substantial semantic gap remains between source code and binary code. 
Source code preserves rich information—such as variable names and types, and structural cues—whereas stripped binaries lose most of these details, rendering direct application of LLMs to such binaries ineffective.

To bridge this gap, we propose a hierarchical summarization–driven domain adaptation strategy, illustrated in Figure~\ref{fig:arch}. We define a code snippet as a function and its constituent statements, where statements correspond to control-flow constructs (e.g., \texttt{if-else}, \texttt{while}, and \texttt{switch}). This strategy employs summaries at two complementary levels to provide a complete and nuanced representation of binary function semantics: function-level summaries capture high-level functional intent, while statement-level summaries model fine-grained implementation logic. Specifically, we extract statements from the abstract syntax tree (AST) and employ an advanced LLM to generate summaries for both statements and entire functions, forming a hierarchical representation that are used to systematically enhances \textsc{Hieronym}’s understanding of binary semantics.

\noindent \textbf{Statement-Level Data Preparation.}
Compared with source code, binaries with symbol tables are structurally closer to their stripped counterparts while preserving richer information, such as identifier names and type details. Leveraging this property, we parse decompiled functions from binaries with symbol tables using Tree-sitter~\cite{tree-sitter} to construct ASTs. From the ASTs, we identify nodes corresponding to control-flow constructs—such as \texttt{if-else}, \texttt{for}, \texttt{while}, \texttt{do-while}, and \texttt{switch}—as candidate statements, and further decompose nested constructs into multiple independent candidate statements.
To improve statement quality, we apply a two-stage refinement process. First, we merge overly short fragments (those containing fewer than five lines of meaningful code) to reduce noise. Second, we merge overlapping fragments whose line-count difference is less than eight lines to eliminate redundancy. The resulting set consists of moderate-sized statements, which serve as the fundamental units for subsequent hierarchical summary generation.

\noindent \textbf{Summary Generation via Code LLMs.}
Building on the success of code-specialized LLMs (such as Qwen3-Coder~\cite{qwen3technicalreport} and Code Llama~\cite{rozière2024codellamaopenfoundation}) in generating high-quality function summaries, we leverage these models to guide binary code analysis. 
We do not use original source-code comments, as prior studies show that LLM-generated summaries yield significantly better performance than source-code comments for function renaming~\cite{symgen}.
Motivated by these findings, we employ Qwen3-Coder to generate summaries of code snippet from decompiled binaries that retain symbol tables. These summaries provide high-quality semantic supervision for adapting \textsc{Hieronym} to the binary domain.
During summary generation, we guide the LLM to focus on stable features (e.g., inputs/outputs, key steps, and logical structure), ensuring closer alignment with the characteristics of stripped decompiled code.

To improve reliability, we employ self-consistency checking~\cite{wang-etal-2023-self-instruct, chen2024universal, min2024beyond}: multiple candidate summaries are generated through repeated LLM queries, the model then evaluates their mutual consistency, and the most consistent summary is selected as the final output. Through experimentation with different prompting strategies, we identify an optimal prompt template (shown in Figures~\ref{fig:prompt1} and~\ref{fig:prompt2} in Appendix~\ref{code_summary}).

\noindent \textbf{Alignment of Stripped Binaries and Summaries.}
To associate summaries derived from DWARF-enabled binaries with stripped binaries, we perform alignment at both the function and statement levels, constructing (stripped function/statement, summary) pairs for training the binary-adapted LLM, as shown in Figure~\ref{fig:arch}.
We first conduct function-level alignment by parsing DWARF information and matching stripped functions to their unstripped counterparts based on function addresses (i.e., start and end addresses). As symbol-table stripping does not modify function address, the address-based matching establishes a reliable correspondence, enabling accurate association of summaries with stripped functions.

We then perform statement-level alignment. The stripping process may alter internal function structure, so the line-number-based matching is unreliable. To address this issue, we propose a hybrid alignment strategy that combines structural and semantic similarity to match statements between stripped and unstripped binaries, making the alignment robust to local variations introduced by stripping. Specifically, we parse decompiled code using Tree-sitter to identify logical statements (e.g., \texttt{if-else}, \texttt{for}, \texttt{while}). For each statement, we construct its AST and extract type paths from the root to leaf nodes. Structural similarity is measured using the Jaccard similarity over these AST-path sets. 
We obtain semantic embeddings for each statement using a pre-trained CodeBERT model~\cite{feng-etal-2020-codebert} and compute semantic similarity via cosine similarity. The matching score is defined as a weighted combination of structural and semantic similarities.
During alignment, we enforce consistency in statement types and accept a match only if its score exceeds a predefined threshold.
This procedure enables mapping of stripped statements to summaries generated from their unstripped counterparts.

Based on the above design, we adopt hierarchical summarization as our domain-adaptation strategy. 
Specifically, we pair decompiled stripped binary code snippets with the corresponding summaries and further fine-tune \textsc{Hieronym} to adapt it to binary semantics.

\subsection{Multi‑Source Information Fusion}
To acquire task‑relevant semantics, we introduce a semantic fusion module that integrates multi-source information to construct a comprehensive representation of function semantics. Specifically, the module incorporates three complementary sources: global binary context, local calling context, and the target function itself. This section describes the representation of each information.

\noindent\textbf{Global Binary Context.}
Motivated by reverse engineering practice, we extract domain-relevant information from binaries and denote it as global binary context.
String literals—which may remain present even in stripped binaries—serve as a primary source of domain-relevant information. These strings are generally stable across compilation settings and often encode domain- or function-related semantic cues. We extract candidate strings by retaining sequences longer than eight characters and applying heuristic filters to remove meaningless patterns.
Specifically, we discard low-quality strings (e.g., low-diversity or non-alphanumeric–dominated patterns) and assign each remaining string a quality score in [0,1] based on alphanumeric ratio, proportion of non-special characters, and penalties for consecutive special characters. Only strings with scores above 0.65 are retained, while system-related strings (e.g., compilation paths) are preserved.
From the filtered set, we randomly sample 100 strings and augment them with file-type indicators and dynamic dependency information to construct the final string set.
File information provides insights into architecture and build characteristics, while dynamic dependencies that indicate domains(e.g., libssl/libcurl suggesting encryption/networking), jointly helping to constrain the binary’s domain.
However, string content varies substantially across binaries, and naively concatenating all strings can easily exceed the context-length limits of \textsc{Hieronym} during fine-tuning. To address this issue, we leverage the summarization capability of an LLM to generate a fixed-length textual summary from the final string set. 
This summary captures high-level domain context, yielding a compact yet semantically rich representation of global binary context, as illustrated in Figure~\ref{fig:Example Binary Information} of the Appendix~\ref{code_summary}.

\noindent \textbf{Local Calling Context.}
We construct local calling context based on the direct call relationships of the target function. 
Since call-path depth varies significantly across functions, incorporating all indirect relationships would produce excessively large call graphs, incurring substantial computational and memory overhead during training and inference
To balance expressiveness and efficiency, we follow prior work~\cite{10.1145/2950290.2950350,8835340,symlm} and restrict context to direct callers and callees.
These functions are categorized as internal or external. As illustrated in Figure~\ref{fig:arch}, the direct context of target function (node 3) includes two callers (nodes 1 and 2) and two callees (nodes 4 and 5), where nodes 1–4 are internal functions and node 5 is external.

We adopt distinct representation strategies for internal and external functions. For internal functions, we use function summaries as their representations. Although decompiled code is available, directly incorporating it presents two challenges: (i) its length may exceed the input limits of \textsc{Hieronym}, and (ii) long caller or callee code can dilute attention and hinder the model’s focus on the target function. For example, in Figure~\ref{fig:motivation}, the decompiled code of the caller \textit{png\_read\_row} spans 156 lines, whereas the target function contains only 30 lines.
Function summaries provide a compact and effective alternative. First, summaries are less sensitive to instruction-level variations introduced by compilation, yielding robust representations. Second, compared with name-propagation approaches~\cite{llasm}, which require iterative execution and are prone to error accumulation, summaries convey richer semantics in a concise form and enable parallel processing, improving both efficiency and stability. Finally, summaries reduce input length, better fit the context window of LLMs, and help the model focus on target function semantic.
For external functions whose implementations are unavailable (e.g., dynamically linked library functions), we exploit the observation that their semantics can typically be inferred from their names (e.g., Linux system calls~\cite{Linux}). Accordingly, we directly use function names as their representations.
The constructed local calling context comprises two components: summaries of internal functions and names of external functions. This design captures both internal implementation semantics and external calling interfaces, yielding a semantically rich and robust representation.

\noindent \textbf{Target Function.}
We represent the target function using its decompiled code extracted from stripped binaries. Unlike raw assembly, decompiled code preserves richer structural and semantic information, making it more amenable to LLMs. Prior work has shown that using decompiled code as input can improve function renaming performance~\cite{symgen}.
However, decompiler-generated code typically contains address-based function names that carry no semantic meaning. Directly incorporating such names can mislead model learning and degrade semantic understanding~\cite{symgen}. Therefore, we preserve the original code structure—including function bodies and parameters—while replacing address-based function names with a special \texttt{[MASK]} token. This strategy removes artificial naming noise and aligns the input representation with the masked-language modeling paradigm commonly used during LLM pretraining, thereby improving compatibility with the model’s learned representations.

\subsection{Function Renaming Fine‑Tuning}
\textsc{Hieronym} aims to achieve accurate and generalizable function renaming for stripped binaries. To this end, we fine-tune the LLM—already adapted to the binary domain—on a binary function renaming dataset, thereby optimizing its performance for the target task.

\noindent \textbf{Parameter-Efficient Learning.}
Most existing binary function renaming models (e.g., SymLM~\cite{symlm} and BLens~\cite{blen}) rely on full-parameter supervised fine-tuning, which updates all model parameters and incurs substantial computational cost. For example, SymLM requires up to 8 days of fine-tuning~\cite{symlm}. Given that \textsc{Hieronym} is built upon a significantly larger LLM, full-parameter fine-tuning would be prohibitively expensive.
To address this issue, we adopt a parameter-efficient fine-tuning strategy based on Low-Rank Adaptation (LoRA)~\cite{hu2022lora}. LoRA freezes the original model parameters and introduces trainable low-rank matrices, dramatically reducing the number of trainable parameters while often achieving performance comparable to—or better than—full fine-tuning~\cite{hu2022lora,vu2022spotbetterfrozenmodel}.

\textsc{Hieronym} is built on the LLM with a Transformer decoder architecture, whose core layers include masked multi‑head attention, multi‑head attention, additive normalization, and feed‑forward layers.
To efficiently adapt the model to binary semantics, we apply LoRA to the attention layers, which play a central role in semantic representation learning.
Specifically, the original attention weight matrices are frozen to preserve the model’s pre-trained language understanding and instruction-following capabilities.
During training, gradient updates ( \(\delta W \in \mathbb{R}^{d \times d}\)) are instead captured through a low-rank decomposition into two trainable matrices \(W_\nabla \in \mathbb{R}^{d \times r}\) and \(W_\Delta \in \mathbb{R}^{r \times d}\). This reduces trainable parameters from \(d \times d\) to \(2 \times r \times d\) ($r \ll d$), yielding substantial savings in computation and memory.
The low-rank updates are incorporated into the original model via residual connections~\cite{he2015deepresiduallearningimage}, preserving expressive capacity while enabling adaptation to the binary function renaming task.

\noindent \textbf{Fine-Tuning Ground Truth and Objective.}
During the fine‑tuning, we follow the common practice established in prior work~\cite{nero,NFRE,symlm,Epitome,symgen} and use the original developer‑defined function names as ground truth which are extracted from binaries with symbol tables.
To ensure compatibility with the model’s output space, each ground-truth name is tokenized using the tokenizer associated with the base model\footnote{We use the Hugging Face tokenizer provided with the base model of \textsc{Hieronym}.}. 

We formulate binary function renaming as an autoregressive sequence generation task. 
The input to \textsc{Hieronym} is denoted as $\mathcal{I}$, which consists of three components: global binary context $\mathcal{B}$, local calling context $\mathcal{C}$, and target function $\mathcal{T}$. 
The calling context $\mathcal{C}$ includes the callers ($\mathcal{C}^+$) and callees ($\mathcal{C}^-$) of $\mathcal{T}$, i.e., $\mathcal{C}=\{\mathcal{C}^+,\mathcal{C}^-\}$, and the input is $\mathcal{I}=\{\mathcal{B},\mathcal{C},\mathcal{T}\}$.
Given a tokenized input sequence \(\mathcal{I} = \{i_1, i_2, \dots, i_n\}\) and an output sequence \(W = \{w_1, w_2, \dots, w_m\}\), \textsc{Hieronym} employs a forward autoregressive factorization and is trained to maximize the conditional log-likelihood:
\begin{align}
    \max_\theta \; \log P_\theta(W) = \sum_{t=1}^{m} \log P_\theta\bigl(w_t \mid \mathcal{I}; \, w_{1:t-1}\bigr) 
\end{align}
\begin{align}
    \log P_{\theta}\bigl(w_t \mid \mathcal{I}; w_{1:t-1}\bigr) 
= \frac{\exp\!\bigl(\mathbf{h}_{\theta}(\mathcal{I}; w_{1:t-1})^{\top} \mathbf{e}(w_t)\bigr)}
{\sum_{w' \in \mathcal{V}} \exp\!\bigl(\mathbf{h}_{\theta}(\mathcal{I}; w_{1:t-1})^{\top} \mathbf{e}(w')\bigr)}
\end{align}

where \(\theta\) denotes the trainable parameters of \textsc{Hieronym}, \(w_{1:t-1}\) represents the sequence of tokens generated prior to predicting \(w_t\), \(\mathbf{h}_{\theta}(\cdot)\) is the hidden state vector produced by the intermediate layers of \textsc{Hieronym} for contextual representation, \(\mathbf{e}(\cdot)\) is the corresponding token embedding, and \(\mathcal{V}\) is the vocabulary of \textsc{Hieronym}.
\section{Evaluation}
In this section, we conduct a comprehensive experimental evaluation to answer the following research questions:

\noindent\textbf{RQ1:} How well does \textsc{Hieronym} rename stripped-binary functions?

\noindent\textbf{RQ2:} How does \textsc{Hieronym} compare with SOTA methods?

\noindent\textbf{RQ3:} How does each component affect overall performance?

\noindent\textbf{RQ4:} How well does \textsc{Hieronym} generalize?

\subsection{Experimental Setup}
\noindent \textbf{Environment.}
All experiments are run on a server equipped with 48-core Intel Xeon Platinum 8558P CPU, 2 TB of RAM, and 8*NVIDIA H100 GPUs, running Ubuntu 22.04 with Python 3.10. We use Ghidra~\cite{Ghidra} (version 11.2.1) to extract functions from binaries and generate decompiled code. Although \textsc{Hieronym} is decompiler-agnostic, Ghidra is selected due to its open-source availability and widespread adoption in prior research.
To ensure fair comparison, we adopt Code Llama model as the base model, consistent with the SOTA method SymGen~\cite{symgen}. Code Llama is a code-specialized variant of Llama 2, pretrained on six programming languages~\cite{rozière2024codellamaopenfoundation}.
Parsing of decompiled code is implemented using ANTLR 4~\cite{parr2013definitive}. Parameter-efficient fine-tuning is conducted using PyTorch~\cite{paszke2019pytorch}, the Hugging Face Transformers library~\cite{wolf2020huggingfacestransformersstateoftheartnatural}, and LlamaFactory~\cite{zheng2024llamafactory}.
For function-name preprocessing and evaluation, we use NLTK~\cite{bird2009natural}, python-Levenshtein~\cite{b27}, and the Qwen3-Embedding model~\cite{qwen3embedding}.

\noindent \textbf{Baselines.}
In this paper, we select SymLM~\cite{symlm}, XFL~\cite{patrick2023xfl}, SymGen~\cite{symgen}, and BLens~\cite{blen} as baselines for comparison. All baselines are trained on our dataset under the same experimental settings as \textsc{Hieronym} to ensure a fair comparison, and we report results from the best-performing hyperparameter configurations following each method’s training guidelines.
XFL is applicable only to x64 binaries and BLens builds upon XFL’s representations, both XFL and BLens are evaluated exclusively on the x64 dataset and compared against \textsc{Hieronym}.
We exclude NERO~\cite{nero}, NFRE~\cite{NFRE}, or Epitome~\cite{Epitome} from comparison, as prior studies have shown that the selected baselines consistently outperform them.
We also omit llasm~\cite{llasm}, as its full source code is unavailable and its results could not be reproduced despite attempts to contact the authors.

\subsection{Dataset Construction} \label{data_setup}
We evaluate \textsc{Hieronym} on the dataset introduced by SymGen~\cite{symgen}, which comprises 33 open-source projects and their corresponding binaries compiled with GCC 9.4.0 across four optimization levels (O0–O3) and four architectures (x86, x64, ARM, and MIPS) (the detailed statistics are provided in Appendix~\ref{symgen_dataset}). The selected projects originate from widely used libraries, which are extensively employed in prior function renaming studies~\cite{symlm, Epitome, NFRE}.

\noindent \textbf{Function Renaming Dataset.}
The dataset contains 9,842 unique binaries and 2,237,915 binary functions in total. 
Unlike prior methods such as SymGen and SymLM, which train separate models for each architecture–optimization pair, \textsc{Hieronym} trains a single model per architecture, thereby reducing training overhead. 
Following prior work~\cite{symlm,symgen,Epitome}, we randomly split the dataset into training, validation, and test sets in an 8:1:1 ratio.

To prevent data leakage~\cite{Hannun2021MeasuringDL}, we split the dataset based on source code: binaries derived from the same source file but compiled with different optimization levels are assigned to the same sets. 
Although function duplication due to code reuse and shared libraries is common in real-world binaries, we follow the stricter deduplication strategy of SymGen to evaluate generalization. Specifically, we remove functions that share the same name, have identical bodies despite different names, or originate from the same source but differ only due to address-dependent callees. Consistent with previous studies~\cite{patrick2023xfl,blen}, we exclude empty functions, functions with meaningless names, and functions automatically generated by Ghidra.

Since \textsc{Hieronym} adopts the same base model and dataset configuration as SymGen, whose study demonstrated that the dataset was not leaked into the pretraining corpus of Code Llama~\cite{rozière2024codellamaopenfoundation}, potential data-contamination concerns are mitigated, ensuring a fair comparison with SOTA approaches.

\noindent \textbf{Hierarchical Domain Adaptation Dataset.}
Following the procedure described in Section~\ref{statement}, we construct the domain-adaptation dataset exclusively from the training set. Specifically, we first generate summaries for code snippets from decompiled code with symbol tables using a code-specialized LLM, and then align these summaries with the corresponding stripped decompiled code snippets.

\subsection{RQ1: Overall Effectiveness}
\begin{table*}[tb]
    \centering
    \vspace{-5pt}
    \newcommand{\uparrowgreen}{\textcolor[rgb]{0,0.85,0}{\uparrow}}
    \caption{Overall Token‑Level Performance of \textsc{Hieronym} and Our Baselines. Note that $\uparrowgreen$ indicates the improvement of \textsc{Hieronym} over the corresponding baseline.}
    \begin{center}
    \resizebox{\columnwidth*2}{!}{
    \begin{tabular}{cc|ccc|ccc|ccc|ccc}
    \toprule
         \multirow{2}{*}{\textbf{Architecture}}&\multirow{2}{*}{\textbf{Model}}&{\textbf{Precision}}&{\textbf{Recall}}&{\textbf{F1-score}}&{\textbf{Precision}}&{\textbf{Recall}}&{\textbf{F1-score}}&{\textbf{Precision}}&{\textbf{Recall}}&{\textbf{F1-score}}&{\textbf{Precision}}&{\textbf{Recall}}&{\textbf{F1-score}}\\
         \cline{3-14}
         {}&{}&\multicolumn{3}{|c|}{O0}&\multicolumn{3}{c|}{O1}&\multicolumn{3}{c|}{O2}&\multicolumn{3}{c}{O3}\\
         \midrule
         \multirow{9}{*}{x64} &{\textsc{Hieronym}}& 0.5262 & 0.5620 & 0.5435 & 0.5373 & 0.5342 & 0.5357 & 0.4991 & 0.5125 & 0.5057 & 0.5117 & 0.5230 & 0.5173 \\
         &{SymGen}&0.4103 & 0.3933 & 0.4016 & 0.4180 & 0.3953 & 0.4064 & 0.4158 & 0.3750 & 0.3943 & 0.4044 & 0.4000 & 0.4022 \\
         &{BLens\textsuperscript{*}}&0.1416 & 0.1031 & 0.1193 & 0.1397 & 0.1092 & 0.1225 & 0.0926 & 0.0602 & 0.0730 & 0.1188 & 0.0755 & 0.0923 \\
         &{XFL\textsuperscript{*}}& 0.1041 & 0.0788 & 0.0897 & 0.1712 & 0.1126 & 0.1359 & 0.1000 & 0.0721 & 0.0838 & 0.1122 & 0.0680 & 0.0847 \\
         &{SymLM}& 0.0376 & 0.0854 & 0.0522 & 0.0979 & 0.0550 & 0.0705 & 0.0749 & 0.0462 & 0.0571 & 0.0922 & 0.0569 & 0.0704 \\
         &{\textsc{Hieronym} vs SymGen}& $(\uparrowgreen28.24\%)$ & $(\uparrowgreen42.90\%)$ & $(\uparrowgreen35.34\%)$ & $(\uparrowgreen28.54\%)$ & $(\uparrowgreen35.14\%)$ & $(\uparrowgreen31.83\%)$ & $(\uparrowgreen20.03\%)$ & $(\uparrowgreen36.67\%)$ & $(\uparrowgreen28.25\%)$ & $(\uparrowgreen26.53\%)$ & $(\uparrowgreen30.74\%)$ & $(\uparrowgreen28.61\%)$ \\
         &{\textsc{Hieronym} vs BLens}& $(\uparrowgreen271.60\%)$ & $(\uparrowgreen445.12\%)$ & $(\uparrowgreen355.59\%)$ & $(\uparrowgreen284.62\%)$ & $(\uparrowgreen389.19\%)$ & $(\uparrowgreen337.35\%)$ & $(\uparrowgreen438.96\%)$ & $(\uparrowgreen751.35\%)$ & $(\uparrowgreen592.75\%)$ & $(\uparrowgreen330.71\%)$ & $(\uparrowgreen592.68\%)$ & $(\uparrowgreen460.42\%)$ \\
         &{\textsc{Hieronym} vs XFL}& $(\uparrowgreen405.46\%)$ & $(\uparrowgreen613.23\%)$ & $(\uparrowgreen505.93\%)$ & $(\uparrowgreen213.85\%)$ & $(\uparrowgreen374.42\%)$ & $(\uparrowgreen294.22\%)$ & $(\uparrowgreen399.08\%)$ & $(\uparrowgreen610.84\%)$ & $(\uparrowgreen503.47\%)$ & $(\uparrowgreen356.04\%)$ & $(\uparrowgreen669.08\%)$ & $(\uparrowgreen510.70\%)$ \\
         &{\textsc{Hieronym} vs SymLM}& $(\uparrowgreen1299.44\%)$ & $(\uparrowgreen558.11\%)$ & $(\uparrowgreen941.22\%)$ & $(\uparrowgreen448.83\%)$ & $(\uparrowgreen871.27\%)$ & $(\uparrowgreen659.93\%)$ & $(\uparrowgreen566.32\%)$ & $(\uparrowgreen1009.34\%)$ & $(\uparrowgreen785.65\%)$ & $(\uparrowgreen454.97\%)$ & $(\uparrowgreen819.11\%)$ & $(\uparrowgreen634.75\%)$ \\
         \hline
         \multirow{5}{*}{x86}&{\textsc{Hieronym}}& 0.5016 & 0.4085 & 0.4503 & 0.5057 & 0.4671 & 0.4856 & 0.4981 & 0.4700 & 0.4836 & 0.5200 & 0.5065 & 0.5131 \\
         &{SymGen}& 0.3656 & 0.3243 & 0.3437 & 0.4049 & 0.3468 & 0.3736 & 0.3825 & 0.3436 & 0.3620 & 0.3981 & 0.3766 & 0.3871 \\
         &{SymLM}& 0.1518 & 0.0660 & 0.0920 & 0.1385 & 0.0674 & 0.0907 & 0.1017 & 0.0736 & 0.0854 & 0.1320 & 0.0903 & 0.1073 \\
         &{\textsc{Hieronym} vs SymGen}& $(\uparrowgreen37.19\%)$ & $(\uparrowgreen25.96\%)$ & $(\uparrowgreen31.00\%)$ & $(\uparrowgreen24.89\%)$ & $(\uparrowgreen34.68\%)$ & $(\uparrowgreen29.98\%)$ & $(\uparrowgreen30.23\%)$ & $(\uparrowgreen36.78\%)$ & $(\uparrowgreen33.60\%)$ & $(\uparrowgreen30.61\%)$ & $(\uparrowgreen34.49\%)$ & $(\uparrowgreen32.56\%)$ \\
         &{\textsc{Hieronym} vs SymLM}& $(\uparrowgreen230.41\%)$ & $(\uparrowgreen518.90\%)$ & $(\uparrowgreen389.41\%)$ & $(\uparrowgreen265.12\%)$ & $(\uparrowgreen592.96\%)$ & $(\uparrowgreen435.40\%)$ & $(\uparrowgreen389.79\%)$ & $(\uparrowgreen538.54\%)$ & $(\uparrowgreen466.31\%)$ & $(\uparrowgreen293.91\%)$ & $(\uparrowgreen460.89\%)$ & $(\uparrowgreen378.22\%)$ \\
         \hline
         \multirow{5}{*}{ARM}&{\textsc{Hieronym}}& 0.3862 & 0.3835 & 0.3848 & 0.4584 & 0.4542 & 0.4563 & 0.4530 & 0.4489 & 0.4510 & 0.4058 & 0.4015 & 0.4036 \\
         &{SymGen}& 0.3144 & 0.3054 & 0.3098 & 0.2807 & 0.2544 & 0.2669 & 0.3192 & 0.3022 & 0.3105 & 0.2879 & 0.2670 & 0.2771 \\
         &{SymLM}& 0.0587 & 0.0263 & 0.0363 & 0.0472 & 0.0349 & 0.0401 & 0.0645 & 0.0394 & 0.0489 & 0.0496 & 0.0213 & 0.0298 \\
         &{\textsc{Hieronym} vs SymGen}& $(\uparrowgreen22.84\%)$ & $(\uparrowgreen25.56\%)$ & $(\uparrowgreen24.22\%)$ & $(\uparrowgreen63.29\%)$ & $(\uparrowgreen78.54\%)$ & $(\uparrowgreen70.96\%)$ & $(\uparrowgreen41.92\%)$ & $(\uparrowgreen48.55\%)$ & $(\uparrowgreen45.24\%)$ & $(\uparrowgreen40.95\%)$ & $(\uparrowgreen50.37\%)$ & $(\uparrowgreen45.66\%)$ \\
         &{\textsc{Hieronym} vs SymLM}& $(\uparrowgreen557.93\%)$ & $(\uparrowgreen1358.06\%)$ & $(\uparrowgreen960.15\%)$ & $(\uparrowgreen871.12\%)$ & $(\uparrowgreen1201.46\%)$ & $(\uparrowgreen1037.85\%)$ & $(\uparrowgreen602.33\%)$ & $(\uparrowgreen1039.40\%)$ & $(\uparrowgreen822.20\%)$ & $(\uparrowgreen718.11\%)$ & $(\uparrowgreen1784.86\%)$ & $(\uparrowgreen1254.42\%)$ \\
         \hline
         \multirow{5}{*}{MIPS}&{\textsc{Hieronym}}& 0.3543 & 0.2650 & 0.3032 & 0.3630 & 0.2891 & 0.3218 & 0.3684 & 0.2809 & 0.3188 & 0.2902 & 0.2502 & 0.2687 \\
         &{SymGen}& 0.1881 & 0.2029 & 0.1952 & 0.1812 & 0.2074 & 0.1935 & 0.1822 & 0.1787 & 0.1804 & 0.1345 & 0.1560 & 0.1445 \\
         &{SymLM}& 0.0198 & 0.0097 & 0.0130 & 0.0256 & 0.0095 & 0.0139 & 0.0185 & 0.0070 & 0.0102 & 0.0128 & 0.0046 & 0.0068 \\
         &{\textsc{Hieronym} vs SymGen}& $(\uparrowgreen88.38\%)$ & $(\uparrowgreen30.61\%)$ & $(\uparrowgreen55.34\%)$ & $(\uparrowgreen100.31\%)$ & $(\uparrowgreen39.39\%)$ & $(\uparrowgreen66.33\%)$ & $(\uparrowgreen102.21\%)$ & $(\uparrowgreen57.21\%)$ & $(\uparrowgreen76.71\%)$ & $(\uparrowgreen115.75\%)$ & $(\uparrowgreen60.40\%)$ & $(\uparrowgreen85.97\%)$ \\
         &{\textsc{Hieronym} vs SymLM}& $(\uparrowgreen1689.62\%)$ & $(\uparrowgreen2631.95\%)$ & $(\uparrowgreen2232.52\%)$ & $(\uparrowgreen1317.81\%)$ & $(\uparrowgreen2943.21\%)$ & $(\uparrowgreen2215.46\%)$ & $(\uparrowgreen1891.52\%)$ & $(\uparrowgreen3913.38\%)$ & $(\uparrowgreen3025.39\%)$ & $(\uparrowgreen2167.03\%)$ & $(\uparrowgreen5339.78\%)$ & $(\uparrowgreen3851.88\%)$ \\
    \bottomrule
    \end{tabular}
    }
    \label{tab:overall}
    \begin{tablenotes}
    \item[*] *XFL targets only x64 binaries by design~\cite{patrick2023xfl}, and BLens~\cite{blen} builds upon XFL. Accordingly, both are evaluated solely on the x64 dataset.
    \end{tablenotes}
    \end{center}
\end{table*}

Our analysis begins with a comprehensive evaluation of \textsc{Hieronym}’s performance. Token-level results across four architectures and four optimization levels are reported in Table~\ref{tab:overall}, while name-level results across the four architectures are summarized in Table~\ref{tab:overall_2}. Overall, \textsc{Hieronym} demonstrates strong performance, though clear variations emerge across different architectures.
At the token level, \textsc{Hieronym} achieves weighted-macro precision, recall, and F1-score of 0.4487, 0.4223, and 0.4339, respectively. 
When examined by architecture, \textsc{Hieronym} performs best on x64 binaries, attaining an average precision of 0.5186, a recall of 0.5329, and an F1-score of 0.5256. Results on x86 binaries are slightly lower, with a precision of 0.5063, a recall of 0.4630, and an F1-score of 0.4832. In contrast, performance declines on ARM and MIPS binaries: \textsc{Hieronym} attains a precision of 0.4258, recall of 0.4220, and F1-score of 0.4239 for ARM, and the corresponding values are 0.3440, 0.2713, and 0.3031 for MIPS.
In the name-level evaluation, the performance trend is largely consistent with that observed at the token level, although minor deviations arise on the ARM architecture. Specifically, \textsc{Hieronym} achieves the highest accuracy on x64 (0.4254), followed by ARM (0.3455) and x86 (0.3220), with performance dropping markedly on MIPS (0.1709).
As name-level evaluation is conducted on a sampled subset of the test set, these results may be affected by the dataset scale. Nevertheless, the overall trends remain consistent with token-level findings, indicating stable model behavior across architectures.

The observed performance gap between x86/x64 and ARM/MIPS architectures is consistent with that of baselines under identical settings. For example, \textsc{Hieronym} achieves approximately 1.24× higher performance on x64 than on ARM, comparable to the 1.38× improvement reported for SymGen. 
We attribute this disparity primarily to architecture-specific factors, particularly the greater complexity of application binary interfaces in ARM and MIPS. Their calling conventions and register usage patterns introduce additional challenges for function renaming. 
Moreover, empirical analysis shows that ARM binaries compiled from the same source often yield longer and more structurally complex decompiled functions than their x64 counterparts, with a higher incidence of variable-type recovery errors. 
These architecture-dependent characteristics, combined with errors introduced during the decompilation process, exacerbate the difficulty of semantic modeling, thereby contributing to the reduced performance of \textsc{Hieronym} on ARM and MIPS architectures.

In addition, we conduct a manual evaluation on 1,600 predictions sampled from the C experiments to assess the reliability of our evaluation framework.
The results show that token-level semantic alignment achieves 89.8\% agreement with human judgment, while name-level evaluation reaches 89.2\%. Moreover, the name-level metric reduces GPT hallucinations by 11.28\%. These findings demonstrate that our evaluation framework is both reliable and closely aligned with human judgment.

\subsection{RQ2: Baseline Comparison} \label{sota_cmp}

To further deepen our analysis, we compare \textsc{Hieronym} with baselines on the same dataset. Token-level and name-level evaluation results are reported in Table~\ref{tab:overall} and Table~\ref{tab:overall_2}, respectively.

\begin{table}[t]
    \centering
    \caption{Name‑Level Accuracy of \textsc{Hieronym} and SymGen.}
    \begin{center}
    \resizebox{0.7\columnwidth}{!}{
    \begin{tabular}{c|cccc}
    \toprule
         \multirow{2}{*}{\textbf{Model}}&\multicolumn{4}{c}{\textbf{Architecture}} \\
         \cline{2-5}
         &{x64}&{x86}&{ARM}&{MIPS}\\
         \midrule
         {\textsc{Hieronym}}& 0.4254 & 0.3220 & 0.3455 & 0.1709 \\
         {SymGen}&0.3331 & 0.2251 & 0.1530 & 0.0765 \\
    \bottomrule
    \end{tabular}
    }
    \label{tab:overall_2}
    \end{center}
\end{table} 

\noindent \textbf{Token-Level Comparison.}
Overall, \textsc{Hieronym} substantially outperforms all baselines, achieving improvements of up to 396.37\%, 688.01\%, and 547.72\%  in precision, recall, and F1-score, respectively.
Traditional ML–based baselines exhibit notably weaker token-level performance on unseen functions. Among the x64-only methods, BLens achieves an average precision of 0.1232, recall of 0.0870, and F1-score of 0.1018, while XFL reports comparable results (precision 0.1219, recall 0.0829, F1-score 0.0985). SymLM performs the weakest overall, with average precision, recall, and F1-score of 0.0702, 0.0433, and 0.0515, respectively, across all architectures and optimization levels. In contrast, \textsc{Hieronym} consistently surpasses these baselines by a wide margin. Compared with SymLM, \textsc{Hieronym} achieves improvements of 860.27\% in precision, 1598.83\% in recall, and 1255.60\% in F1-score. Although XFL outperforms SymLM on x64 binaries, \textsc{Hieronym} still exceeds XFL by 343.61\% in precision, 566.89\% in recall, and 453.58\% in F1-score. Similarly, relative to BLens, \textsc{Hieronym} delivers achieves gains of 331.47\%, 544.59\%, and 436.52\% in precision, recall, and F1-score, respectively.
Notably, the performance of ML–based baselines on our dataset is substantially lower than that reported in their studies.
For example, SymLM reports an average F1-score of 0.73 on its dataset but attains 0.0515 in our evaluation. 
This discrepancy is primarily attributable to data leakage in prior experimental setups, where duplicated functions between training and test sets inflate performance and obscure true generalization. Consistent with this explanation, SymLM reports F1-score below 0.10 on unseen binaries in its study~\cite{symlm}. Similar degradation on unseen functions has also been reported for XFL~\cite{patrick2023xfl} and BLens~\cite{blen}. 
Moreover, baseline methods that incorporate calling information are highly sensitive to optimizations: SymLM’s F1-score varies by 35.06\%, XFL by 62.17\%, and BLens by 67.81\% across optimization levels. \textsc{Hieronym} exhibits only a 7.48\% variation, demonstrating strong robustness to optimization level changes.
A similar trend appears across architectures. SymLM shows F1-score variation up to 753.64\%, while \textsc{Hieronym} exhibits substantially smaller variation (73.41\%), highlighting robustness to architectural differences.

The LLM-based baseline SymGen achieves average token-level precision, recall, and F1-score of 0.3180, 0.3018, and 0.3093 across all architectures and optimization levels. Since we adopt the same configuration as in the original SymGen study~\cite{symgen}, these results are not far from those previously reported.
Even under this strong and directly comparable baseline, \textsc{Hieronym} demonstrates a clear advantage, improving precision by 50.12\%, recall by 41.75\%, and F1-score by 45.10\%. 
To further ensure fairness, we also re-evaluate performance using SymGen’s evaluation metrics, under which \textsc{Hieronym} still improves F1-score by 47.96\% over SymGen (see Appendix~\ref{symgen_eval}).
These results indicate that LLM-based methods generalize more effectively to unseen binaries than traditional ML methods, and that the multi-source information and hierarchical summarization–driven domain adaptation employed by \textsc{Hieronym} further enhance semantic understanding and naming performance.

\noindent \textbf{Name-Level Comparison.}
Due to budget constraints and the fact that ML–based methods are restricted to predicting labels from a closed vocabulary, we perform name-level evaluation only for \textsc{Hieronym} and SymGen, as SymGen substantially outperforms other baselines at the token level. Under this setting, \textsc{Hieronym} achieves an average accuracy improvement of 79.94\% over SymGen across all architectures. This result further demonstrates that \textsc{Hieronym} captures function semantics more comprehensively and generates function names with higher semantic accuracy.

\begin{figure}[tb]
    \centering
    \vspace{-4pt}
    \includegraphics[width=0.75\linewidth]{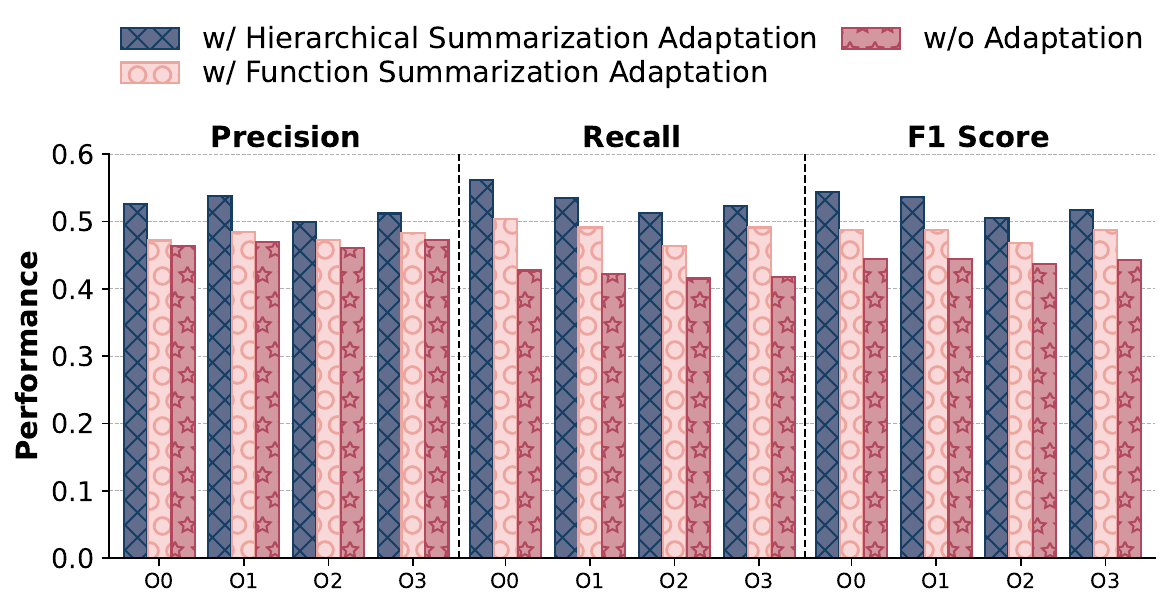}
    \caption{Effectiveness of Hierarchical Domain Adaptation}
    \label{fig:noadop}
\end{figure}

\begin{figure}[tb]
    \centering
    \includegraphics[width=0.75\linewidth]{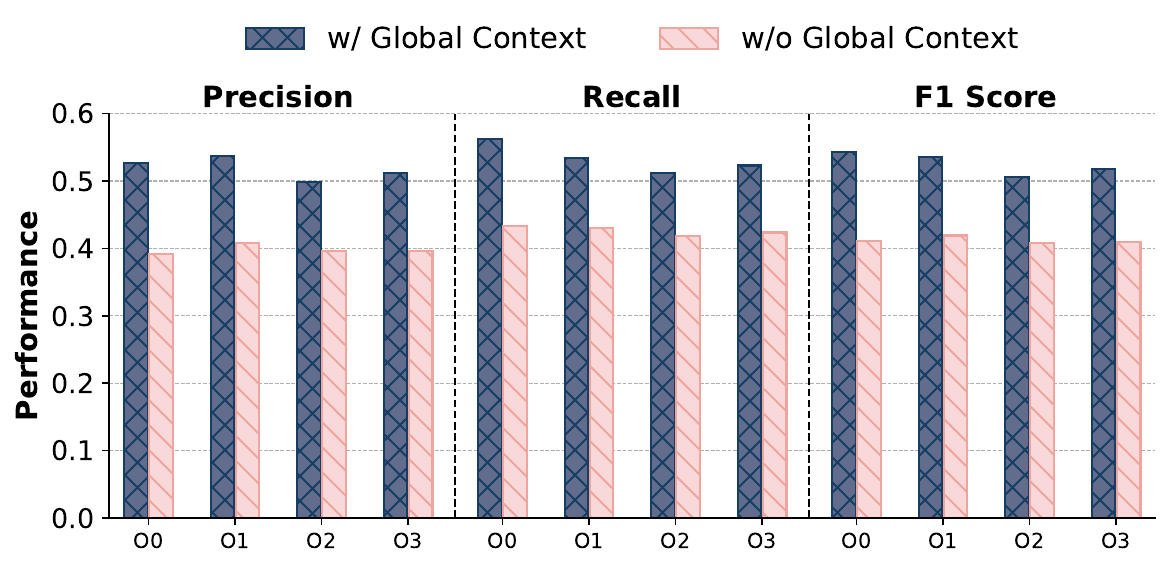}
    \caption{Effectiveness of Global Binary Context}
    \label{fig:nobin}
\end{figure}

\begin{figure}[tb]
    \centering
    \includegraphics[width=0.75\linewidth]{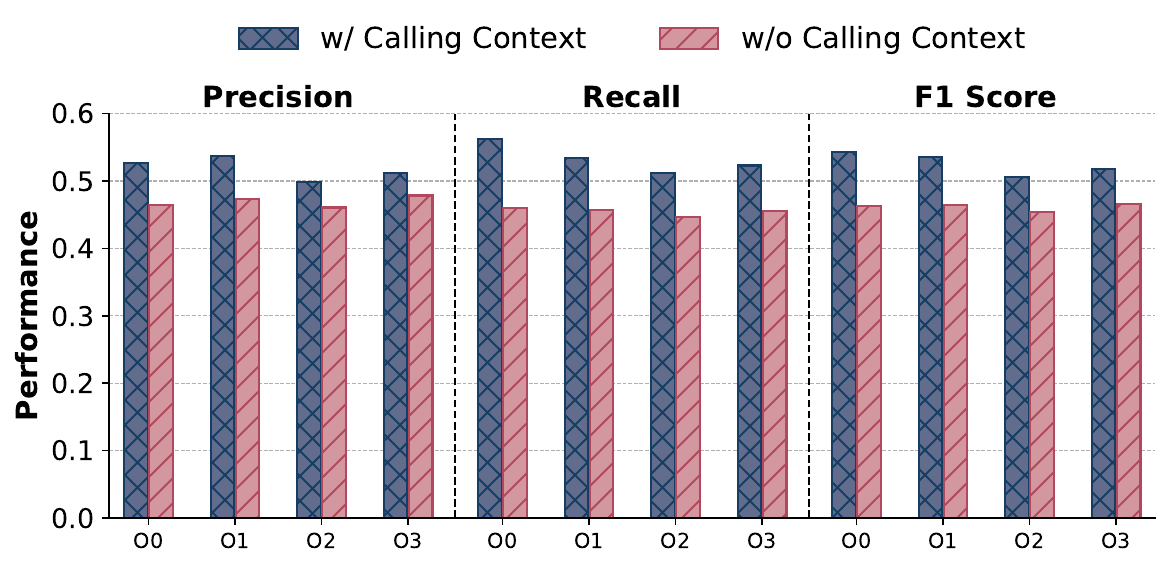}
    \caption{Effectiveness of Calling Context}
    \label{fig:nocall}
\end{figure}

\subsection{RQ3: Ablation Study}
In this section, we conduct ablation studies to obtain deeper insights into \textsc{Hieronym}. Our goal is to systematically evaluate the contribution of each individual component to the overall performance. To ensure a controlled and consistent comparison, all ablation experiments are performed on x64 dataset and use token-level metrics.

\noindent \textbf{Effectiveness of Hierarchical Domain Adaptation.}
Domain adaptation is designed to enhance \textsc{Hieronym}’s ability to capture binary code semantics by bridging the knowledge gap between the general-purpose pre-trained LLM and the binary domain. To evaluate the effectiveness of this strategy, we compare three model configurations: (1) hierarchical summarization-based domain adaptation, (2) function summarization only domain adaptation, and (3) no domain adaptation, where the model is directly fine-tuned on the function-renaming task using parameter-efficient methods. 
This comparison allows us to isolate and quantify the contribution of hierarchical summarization-based domain adaptation.

The results are shown in Figure~\ref{fig:noadop}. Across the four optimization levels (O0–O3) on the x64 architecture, hierarchical summarization-based domain adaptation improves \textsc{Hieronym} over the no-adaptation baseline by an average of 11.14\%, 26.76\%, and 18.83\% in precision, recall, and F1-score, respectively. 
Compared with an LLM adapted using function summarization only adaptation, \textsc{Hieronym} yields average gains of 8.49\%, 9.43\%, and 8.95\% on the same metrics.
Function summarization only adaptation also outperforms the no-adaptation configuration, achieving average improvements of 2.44\%, 15.85\%, and 9.06\% in precision, recall, and F1-score.
These results lead to two key observations. First, domain adaptation enhances the LLM’s ability to understand binary code semantics. Second, incorporating hierarchical summarization provides additional benefits, demonstrating that fine-grained, statement-level semantic learning effectively complements function-level understanding and plays a important role in improving function renaming performance.

\noindent \textbf{Effectiveness of Global Binary Context.}
To assess the impact of global binary context, we compare fine-tuned variants of \textsc{Hieronym} with and without incorporating this context as input. 
As shown in Figure~\ref{fig:nobin}, incorporating global context yields a consistent performance gain, increasing the average F1-score from 0.4117 to 0.5256 (a relative improvement of 27.63\%).
This demonstrates the critical role of global context in function renaming, as it provides domain-relevant information that enhances the model’s ability to capture function semantics and generate accurate names.

To further verify that these gains are not driven by binary names that may appear in the global context, we remove test samples whose context contains such information and re-evaluate on the filtered set. \textsc{Hieronym} achieves an F1-score of 0.444 on this subset, compared to 0.4339 on the original test set. The stable performance indicates that the improvements primarily arise from effective use of domain-relevant context rather than leakage from binary names.

\noindent \textbf{Effectiveness of Calling Context.}
To evaluate the contribution of calling-context information, we compare two fine-tuned variants of \textsc{Hieronym}: one that incorporates calling context in the input and another that excludes it. As shown in Figure~\ref{fig:nocall}, incorporating calling context consistently improves performance, yielding average gains of 10.52\% in precision, 17.22\% in recall, and 13.81\% in F1-score.
Notably, these improvements are more pronounced at lower optimization levels (O0/O1) and diminish at higher levels (O2/O3).
We attribute this trend to the effects of aggressive compiler optimizations, such as function inlining, which remove explicit inter-function call relationships, thereby reducing the effectiveness of calling context.
Overall, these results demonstrate that calling context provides valuable complementary semantic cues, enabling \textsc{Hieronym} to construct richer function semantics representations and improve function renaming performance.

\subsection{RQ4: Generalizability Test}
\begin{table}[t]
\centering
\caption{Cross-Compiler Performance}
\label{tab:cross_compiler}
\resizebox{\columnwidth}{!}{
\begin{tabular}{c|ccc|ccc}
\toprule
\multirow{2}{*}{\textbf{Architecture}}& \multicolumn{3}{c|}{\textbf{\textsc{Hieronym}}} & \multicolumn{3}{c}{\textbf{SymGen}} \\
\cline{2-7}
{}&{\textbf{Precision}}&{\textbf{Recall}}&{\textbf{F1-score}}&{\textbf{Precision}}&{\textbf{Recall}}&{\textbf{F1-score}}\\
\midrule
x64 & 0.5841& 0.6240 &0.6034&0.4447 &0.3950 &0.4178   \\
x86& 0.5462& 0.4916 &0.5175 & 0.4253 &0.3780 &0.4001  \\
ARM& 0.3848 &0.3442 &0.3633 & 0.2708 &0.2346 &0.2512  \\
MIPS& 0.3626 &0.3396 &0.3507  & 0.1966 &0.2038 &0.2000 \\
\bottomrule
\end{tabular}
}
\end{table}

In this section, we conduct a comprehensive evaluation of \textsc{Hieronym}’s generalization across multiple dimensions, including cross-compiler, cross-decompiler, cross-project, and obfuscated settings. We use SymGen as the sole baseline, as it achieves the best performance in Section~\ref{sota_cmp}. To ensure fair comparison, all experiments adopt the same test set split and strict deduplication strategy as in Section~\ref{data_setup}, and are evaluated using token-level metrics.

\noindent \textbf{Cross-Compiler Evaluation.}
We recompile the 33 projects in SymGen dataset using Clang 14.0.0 (details in Appendix~\ref{symgen_dataset}) and re-evaluate both \textsc{Hieronym} and SymGen on the resulting binaries. As shown in Table~\ref{tab:cross_compiler}, \textsc{Hieronym} achieves a weighted macro F1-scores of 0.4587, compared to 0.3173 for SymGen, corresponding to a 49.68\% improvement (see Appendix~\ref{cross_com_result} for details). 
On the GCC-compiled test set, the F1-scores are 0.4339 for \textsc{Hieronym} and 0.3093 for SymGen. The comparable performance across compilers indicates strong cross-compiler generalization for both methods, likely because LLM-based methods operate on decompiled code, which mitigates compiler-specific variation. Notably, \textsc{Hieronym} further benefits from its design, leading to improved naming performance.

\noindent \textbf{Cross-Decompiler Evaluation.}
We re-decompile the test set using IDA Pro 9.0 and re-evaluate both \textsc{Hieronym} and SymGen on the resulting dataset. As shown in Table~\ref{tab:cross_decompiler}, \textsc{Hieronym} achieves a weighted macro-average F1 score of 0.4633, compared to 0.3853 for SymGen, yielding a 20.54\% improvement (see Appendix~\ref{cross_decom_result} for details). 
Compared to Ghidra, IDA can recover additional referenced string information—particularly for ARM and MIPS binaries—which can improve renaming performance. Consequently, both methods benefit from this change, with SymGen showing a more pronounced gain. Nevertheless, \textsc{Hieronym} consistently outperforms SymGen, demonstrating robust cross-decompiler generalization.

\begin{table}[t]
\centering
\caption{Cross-Decompiler Performance}
\label{tab:cross_decompiler}
\resizebox{\columnwidth}{!}{
\begin{tabular}{c|ccc|ccc}
\toprule
\multirow{2}{*}{\textbf{Architecture}}& \multicolumn{3}{c|}{\textbf{\textsc{Hieronym}}} & \multicolumn{3}{c}{\textbf{SymGen}} \\
\cline{2-7}
{}&{\textbf{Precision}}&{\textbf{Recall}}&{\textbf{F1-score}}&{\textbf{Precision}}&{\textbf{Recall}}&{\textbf{F1-score}}\\
\midrule
x64 & 0.4817&0.4793 &0.4805 & 0.4238 &0.3394 &0.3769  \\
x86&  0.4849&0.3994 &0.4379 & 0.4451 &0.3526 &0.3931   \\
ARM& 0.5049 &0.4427 &0.4717  & 0.4167 &0.3573 &0.3846  \\
MIPS&0.4753 & 0.4518 & 0.4632 & 0.4020 &0.3736 &0.3866 \\
\bottomrule
\end{tabular}
}
\end{table}

\noindent \textbf{Cross-Project Evaluation.}
We further deduplicate the test set by removing samples from projects present in the training set and re-evaluate on the resulting dataset.
\textsc{Hieronym} achieves a weighted macro-average precision of 0.4897, recall of 0.4461, and F1-score of 0.4661, while SymGen attains 0.3453 precision, 0.3318 recall, and 0.3377 F1-score.
These results are comparable to those on the original test set, indicating that both methods maintain strong performance on samples from previously unseen projects and exhibit robust cross-project generalization. This also confirms that our deduplication strategy is both strict and effective.

\noindent \textbf{Obfuscated Binary Evaluation.}
To evaluate the obfuscation resistance performance of \textsc{Hieronym}, we use SymGen’s obfuscated binary dataset, which contains 363 binaries from six GNU projects (findutils, coreutils, curl, less, putty, and bash). 
The binaries are obfuscated using three techniques: bogus control flow (bcfobf), control-flow flattening (cffobf), and instruction substitution (subobf).
We also include the original binaries (w/o obfuscation) for comparison. As shown in Table~\ref{tab:obfuscation_performance}, \textsc{Hieronym} achieves F1-scores of 0.3866, 0.3759, and 0.303 across the three obfuscation settings, outperforming SymGen by up to 40.42\%. Relative to original binaries, \textsc{Hieronym} exhibits only modest degradation (e.g., 0.17\% on subobf), which is lower than that of SymGen. These results demonstrate the superior obfuscation robustness of \textsc{Hieronym}, likely due to its generalizable learning paradigm.

\begin{table}[tb]
    \centering
    \caption{Performance on Obfuscated Binaries}
    \label{tab:obfuscation_performance}
    \resizebox{0.7\columnwidth}{!}{
    \begin{tabular}{lcc}
    \toprule
    \textbf{Obfuscation} & \textbf{\textsc{Hieronym}} & \textbf{SymGen} \\
    \midrule
    w/o obfuscation & 0.4095  & 0.2947  \\
    bcfobf          & 0.3866  (-5.58\%)  & 0.2759 (-6.37\%)  \\
    cffobf          & 0.3759 (-8.20\%) & 0.2679 (-9.12\%)  \\
    subobf          & 0.303 (-0.17\%)  & 0.2911 (-1.22\%) \\
    \bottomrule
    \end{tabular}
    }
\end{table}

\noindent \textbf{Cross-Language Evaluation.}
We evaluate the transferability of \textsc{Hieronym} on C++ binaries and Java code. First, we assess \textsc{Hieronym} on the Xalan benchmark (x86-64 architecture) from SPEC CPU2017~\cite{Xalan}, which contains numerous functions with object-oriented C++ characteristics. \textsc{Hieronym} achieves an F1-score of 0.4093, compared to 0.3696 for SymGen. Although both methods perform worse than on the original x64 dataset (0.5256 and 0.4011, respectively), \textsc{Hieronym} still surpasses SymGen by 10.71\%. 
Next, we randomly select four Java projects from AndroZooOpen~\cite{liu2020androzooopen} and apply R8~\cite{R8} to obfuscate identifier names, yielding more than 4,400 functions. We then predict only method names. SymGen achieves an F1-score of 0.2977, whereas \textsc{Hieronym} attains 0.3779, representing a 26.93\% improvement. Compared with C++, performance degrades more substantially on Java, likely due to the larger semantic or structural differences between Java and C-family binaries, which pose additional challenges for generalization. Nevertheless, \textsc{Hieronym} consistently outperforms prior methods across languages, demonstrating strong cross-language transferability.

\section{Illustration}

\begin{table*}[t]
    \setlength{\tabcolsep}{2pt}
    \newcolumntype{P}[1]{>{\raggedright\arraybackslash}p{#1}}
    \caption{Qualitative Examples of Predictions Made by \textsc{Hieronym} and SymGen. \textbf{Tok. (Ours)} denotes our tokenization, and \textbf{Tok. (SymGen)} denotes SymGen's tokenization.}
    \centering
    \footnotesize
    \resizebox{\columnwidth*2}{!}{
    \begin{tabular}{P{1.9cm} P{1.9cm} P{2.2cm}|P{2.3cm} P{2.3cm} P{2cm}|P{1.5cm} P{1.6cm} P{1.7cm}}
    \toprule
    \multicolumn{3}{c|}{\textbf{Ground Truth}}&\multicolumn{3}{c|}{\textbf{\textsc{Hieronym} Prediction}}&\multicolumn{3}{c}{\textbf{SymGen Prediction}}\\
    \hline
    \textbf{Raw} & \textbf{Tok. (Ours)} & \textbf{Tok. (SymGen)} & \textbf{Raw} & \textbf{Tok. (Ours)} & \textbf{Tok. (SymGen)} &\textbf{Raw} & \textbf{Tok. (Ours)} & \textbf{Tok. (SymGen)}\\
    \midrule
        jitterc\_mangle&jitter,c,mangle&jitter,c,mangle&jitter\_mangle&jitter,mangle&jitter,mangle&mangle\_name&mangle,name&mangle,name \\
        collector\_gettid&collector,\allowbreak get, \allowbreak thread, \allowbreak id&collector,\allowbreak get,\allowbreak t,\allowbreak id&collector\_get\_thread\_id&collector,\allowbreak get,\allowbreak thread,\allowbreak id&collector,\allowbreak get,\allowbreak thread,\allowbreak id&get\_thread\_id&get,\allowbreak thread,\allowbreak id&get,\allowbreak thread,\allowbreak id\\
        \hline
        ctf\_type\_encoding&ctf,type,encoding&ct,f,type,encode&ctf\_type\_encoding&ctf,type,encoding&ct,f,type,encode&get\_size&get,size&get,size\\
        check\_message&check,message&check,message&validate\_message&validate,message&validate,message&nc\_mvcur&nc,mv,cur&nc,mvc,ur\\
        \hline
         is\_ascii\_string& is,ascii,string & be,ascii,string&libgettextpo\_is\_ascii\_string&lib,\allowbreak get,\allowbreak text,\allowbreak po,\allowbreak is,\allowbreak ascii,\allowbreak string& lib,\allowbreak get,\allowbreak text,\allowbreak po,\allowbreak is,\allowbreak ascii,\allowbreak string& c\_isascii\_string&c,is,ascii,string& c,\allowbreak isa,\allowbreak sc,\allowbreak ii,\allowbreak string\\
         png\_set\_IHDR&png,set,ihdr&png,set,i,h,d,r&libtextstyle\_png\_set\_ihdr&lib,\allowbreak text,\allowbreak style,\allowbreak png,\allowbreak set,\allowbreak ihdr&lib,\allowbreak text,\allowbreak style,\allowbreak png,\allowbreak set,\allowbreak i,h,d,r& set\_field\_type& set,field,type&set,field,type\\
         \hline
         linearization\_string& linearization,\allowbreak string& linear,\allowbreak ization,\allowbreak string&sensor\_reading\_type\_code\_linearization& {sensor,\allowbreak reading,\allowbreak type,\allowbreak code,\allowbreak linearization}&{sensor,\allowbreak read,\allowbreak type,\allowbreak code,\allowbreak linear,\allowbreak zation}&gnuplot\_scale\_type&gnuplot,\allowbreak scale,\allowbreak type&gnuplot,\allowbreak scale,\allowbreak type\\
         test\_standard\_methods &test,\allowbreak standard,\allowbreak methods& t,\allowbreak est,\allowbreak standard,\allowbreak method&test\_asn1\_standard\_methods&test,\allowbreak asn,\allowbreak standard,\allowbreak methods&t,est,\allowbreak asn,\allowbreak standard,\allowbreak method&test\_pkey\_id\_order&test,\allowbreak pkey,\allowbreak id,\allowbreak order&t,\allowbreak est,\allowbreak pkey,\allowbreak id,\allowbreak order\\
         \bottomrule
    \end{tabular}
    }
    \label{tab:case_study1}
\end{table*}

\subsection{Qualitative Evaluation}
In our qualitative analysis, we examine function names predicted by \textsc{Hieronym} and SymGen, sampled from the inference results. Table~\ref{tab:case_study1} reports both ground-truth and predictions, tokenized using our tokenization method and SymGen’s tokenization method, respectively. 
Tokenization is necessary for token-level evaluation.

Several insightful observations can be drawn from these examples from projects such as poke, binutils, gettext, libpng, freeipmi, and openssl.
First, \textsc{Hieronym} accurately infers function names that include project-specific tokens (e.g., \texttt{jitter\_mangle}). As discussed earlier, although SymGen captures the function’s semantics, it fails to recover such identifiers (e.g., \texttt{jitter}), highlighting the effectiveness of our multi-source information fusion strategy.
Second, \textsc{Hieronym} correctly infers names (e.g., \texttt{ctf\_type\_encoding}) where SymGen fails to capture the correct semantics and produces erroneous predictions. Even when the predictions differ lexically from ground truth, \textsc{Hieronym} often generates semantically equivalent tokens (e.g., {\texttt{check}, \texttt{validate}}), demonstrating a deeper understanding of function semantics. In contrast, SymGen frequently fails to grasp the underlying semantics, yielding less meaningful names.

We also observe a form of inference bias in \textsc{Hieronym}’s predictions, where additional tokens appear that are absent from the ground truth. This phenomenon primarily stems from two sources. (i) Global binary context may include binary names or related library identifiers, which can be incorporated into predictions—for example, tokens such as \texttt{libgettextpo} and \texttt{libtextstyle} reflect domain cues present in the global context. (ii) Calling context may introduce keywords relevant to function behavior but not part of the canonical name. For instance, when predicting \texttt{linearization\_string}, \textsc{Hieronym} introduces tokens such as \texttt{sensor} and \texttt{type} derived from its calling context. 
Despite these extraneous tokens, the predicted names consistently capture the core functionality of the target functions, indicating that \textsc{Hieronym} remains robust in modeling essential function semantics.

Based on the tokenized outputs, we further analyze the differences between our tokenization strategy and SymGen’s, as well as their impact on token-level evaluation results. SymGen often produces overly fine-grained and sometimes erroneous segmentations (e.g., tokenizing `\texttt{test}' into \{\texttt{t}, \texttt{est}\}, `\texttt{ctf}' into \{\texttt{ct}, \texttt{f}\}, or decomposing `\texttt{isascii}' into \{\texttt{isa}, \texttt{sc}, \texttt{ii}\}). In contrast, our approach aims to partition function names into semantically meaningful tokens: `\texttt{isascii}' is correctly tokenized as \{\texttt{is}, \texttt{ascii}\}, while common abbreviations such as '\texttt{ctf}' and frequent words like `\texttt{test}' are retained as single tokens to avoid unnecessary fragmentation.
These differences have a direct and substantial impact on token-level evaluation. Improper tokenization disrupts semantic coherence of tokens and introduces bias into metric computation. For example, for the function name `\texttt{c\_isascii\_string}', the F1-score under SymGen’s tokenization is only 0.25, whereas it increases to 0.857 when using our tokenization strategy. 
This improvement highlights the effectiveness of our approach in preserving semantic integrity during evaluation.
Furthermore, we also conduct name-level evaluation, which is inherently insensitive to tokenization and provides a robust evaluation of the performance of function-renaming methods.

\subsection{Effectiveness on Real-world Binaries}

To evaluate \textsc{Hieronym}'s generalization in real-world security analysis, we apply it to the task of function renaming on malware executables. 
We collect ELF-format malware samples from VirusShare~\cite{Virus-share} and retain only those containing symbol tables to ensure reliable ground-truth function names.
We select various types of x64 Linux malware, such as CDorked, camelot, and DnsAmp. All symbol tables are stripped prior to evaluation, forming an external test set.

Without additional fine-tuning, we directly apply \textsc{Hieronym} trained on our primary dataset—composed mainly of GNU projects—to rename 2,478 malware functions. Evaluation follows token-level metrics described in Section~\ref{sec:evaluation}, comparing predicted names against ground truth extracted from the original symbol tables. \textsc{Hieronym} achieves a precision of 0.4855, a recall of 0.4897, and an F1-score of 0.4876.
Notably, despite substantial differences in functional intent between benign open-source software and malware, performance on malware binaries is comparable to that observed on our evaluation dataset. These results indicate that \textsc{Hieronym} retains robust semantic modeling and name inference capability when applied to previously unseen and domain-divergent binaries, highlighting its strong generalization potential in practical security analysis.

\section{Limitation \& Future Work}

\noindent \textbf{Base Model.} 
To ensure fair experimental comparison, we adopt the same base model and experimental setup as the SOTA method SymGen~\cite{symgen}. However, the landscape of generative LLMs is evolving rapidly, with increasingly capable models continually emerging. As \textsc{Hieronym} is model-agnostic by design, its performance could be further improved by integrating larger or more advanced base models. 
While our experiments already demonstrate the effectiveness of the proposed approach, investigating its integration with alternative base models represents a promising direction.

\noindent \textbf{Statement Extraction Effectiveness.}
During dataset construction, we employ heuristic rules to extract code statements from functions. The lack of objective metrics for assessing statement quality introduces potential limitations and leaves room for further refinement. 
Despite this, our experimental results demonstrate that statement-level summarization effectively supports binary function renaming. Future work will focus on improving the statement extraction algorithm to further enhance performance.

\noindent \textbf{Dataset Quality.}
Dataset quality is a critical factor in effective model training. For the function renaming task, we employ Ghidra to generate decompiled code as model input. Nevertheless, decompiler outputs inevitably contain inaccuracies due to the inherent challenges of decompilation, and discrepancies across different decompilers may introduce systematic bias. Although \textsc{Hieronym} is designed to be decompiler-agnostic, its performance could benefit from higher-fidelity decompilation results.
In addition, the quality of function names directly affects dataset validity. 
Many non-descriptive or semantically ambiguous names remain prevalent in real-world binaries.
Fundamentally addressing this noise typically requires expert-driven manual curation, which is costly and time-consuming. Developing automated techniques to filter low-quality function names represents an important direction for future work.

For the domain-adaptation dataset, we employ Qwen3-Coder-30B to generate natural-language summaries of code snippets, which serve as supervision for training \textsc{Hieronym}. The quality and stylistic consistency of these summaries may vary depending on the LLM used. 
Although proprietary models (e.g., ChatGPT) often produce higher-quality outputs, their cost renders them impractical for large-scale dataset construction.
Prior work~\cite{symgen} has shown that summaries generated by open-source LLMs can still effectively support function renaming task. We therefore expect that further improving the consistency of code-snippet summaries could yield additional performance gains for \textsc{Hieronym}. Exploring this direction constitutes a promising direction for future research.

\noindent \textbf{Evaluator Selection.} 
Using local LLMs similar to those in \textsc{Hieronym} as evaluators may introduce bias. We therefore adopt GPT as a widely used external judge. We further evaluate Qwen3-Code-30B on 1,600 samples from the C experiments as an alternative evaluator. It achieves 76.31\% agreement with human judgments, substantially lower than GPT’s 87.56\%. Developing open-source evaluators that match GPT’s performance is an important direction.

\noindent \textbf{Inference Efficiency and Cost.}
\textsc{Hieronym} requires an average of 20.69 tokens and 0.0333s per function name, compared to 14.86 tokens and 0.0245s for SymGen. \textsc{Hieronym}'s total cost comprises global-context summarization (0.0371s), calling-context summarization (0.0351s), and name generation (0.0333s), with the first two stages parallelizable. On large binaries (2,284 functions), \textsc{Hieronym} achieves a throughput of 3.27 functions/s, corresponding to an end-to-end inference latency of 0.3056s per function.

\section{Related Work}
\subsection{Function Renaming For Stripped Binaries}
Early research on binary function renaming primarily relied on traditional ML techniques. For example, Debin~\cite{Debin} transforms binary code into an intermediate representation and constructs variable dependency graphs, using Conditional Random Fields (CRFs) to infer function names. Similarly, Punstrip~\cite{Prob_name} generates probabilistic binary fingerprints from engineered features and applies a CRF-based model for name prediction. XFL~\cite{patrick2023xfl} formulates function renaming as a multi-label classification problem that predicts sets of words.
More recent works have framed function renaming as a neural machine translation task, translating binary code into natural-language function names. NERO~\cite{nero} represents calling context using augmented control-flow graphs (CFGs) derived from assembly instructions, but largely ignores non-call instructions, which are unavailable for many functions. NFRE~\cite{NFRE} proposes structure-sensitive embeddings of assembly instructions to generate function names, while AsmDepictor~\cite{AsmDepictor} employs a Transformer-based model to encode function semantics from assembly code.
SymLM~\cite{symlm} leverages a pretrained assembly-language model to capture execution behavior, and Epitome~\cite{Epitome} combines a pretrained assembly-language model with graph neural networks over CFGs to learn function semantics. BLens~\cite{blen} adopts a different perspective by casting function renaming as an image-captioning-like task, aggregating multiple assembly-based embeddings into an ensemble representation aligned with the latent space of function names.
Despite their methodological differences, these approaches share a fundamental limitation: function names are typically generated by selecting tokens from a closed training vocabulary, which restricts generalization to unseen or semantically novel names.
Recently, LLMs have demonstrated strong generalization capabilities for function renaming. llasm~\cite{llasm} integrates a pretrained assembly encoder with a decoder-only LLM for name generation. 
SymGen~\cite{symgen} uses decompiled code as its input and introduces a function-summarization-based adaptation strategy to guide name generation.
In this work, we discuss the limitations of existing methods (Section~\ref{motivation}) and compare \textsc{Hieronym} with four SOTA methods—SymLM, XFL, BLens, and SymGen—in Section~\ref{sota_cmp}.

In addition, with the rapid advancement of LLMs, researchers have explored their applications to other stripped binary analysis tasks, including decompilation~\cite{10.1145/3728958,11334626}, variable name recovery~\cite{xie2024resym,xu2025unleashing}, and variable type recovery~\cite{xie2024resym,wang2025typeforge,dramkoidioms}. 
LLM-based approaches introduce a new paradigm for stripped binary analysis and continue to drive progress in the field. 
These efforts complement our work and further advance research on stripped binary analysis.

\subsection{Source Code Summarization}
Source code summarization aims to generate concise and human-readable natural-language descriptions for code snippets. Since around 2016, the majority of approaches in this area have been data-driven~\cite{b15}. One of the earliest seminal works~\cite{b15} introduced an attention-based encoder-decoder architecture inspired by neural machine translation. Building on this paradigm, CodeBERT~\cite{feng-etal-2020-codebert}, a Transformer-based model pre-trained on paired programming and natural-language data, demonstrated strong performance when fine-tuned for statement-level summarization. 
Subsequent research has focused on exploiting richer structural information from source code. Several studies adopted graph-based neural networks to encode AST structures~\cite{b20,b21}. Huang et al.~\cite{HUANG2020106373} further incorporated reinforcement learning over both code tokens and AST sequences to improve statement-level summarization. 
With the recent success of LLMs in natural language processing, increasing attention has been directed toward leveraging them for source code summarization. Sun et al.~\cite{sun2023automaticcodesummarizationchatgpt} demonstrated that LLMs can generate high-quality code summaries using zero-shot prompting. Subsequent studies~\cite{ahmed2022fewshottrainingllmsprojectspecific,10.1145/3597503.3608134} showed that few-shot prompting can significantly outperform supervised approaches. These results highlight the strong potential of LLMs to produce accurate and semantically rich code summaries.
In this work, we leverage these advances to construct a domain-adaptation dataset that provides high-quality semantic supervision for binary code comprehension.

\section{Conclusion}
We propose \textsc{Hieronym}, a novel approach that leverages the strong generalization capabilities of LLMs to rename functions from decompiled code in stripped binaries. To bridge the knowledge gap between general-purpose LLMs and the binary domain, we introduce a hierarchical summarization-driven domain adaptation strategy. \textsc{Hieronym} further integrates multi-source information to construct comprehensive, task-relevant semantics of binary functions, thereby substantially improving function-renaming performance. In addition, we propose a dual-level evaluation framework that combines token-level and name-level metrics to enable systematic and reliable evaluation.
Experimental results show that \textsc{Hieronym} improves token-level precision, recall, and F1-score by 50.12\%, 41.75\%, and 45.10\%, respectively, and achieves a 79.94\% improvement in name-level accuracy over the SOTA method. 
Ablation studies confirm the effectiveness of key design components, and experiments on real-world malware binaries further demonstrate \textsc{Hieronym}’s practical utility in security analysis.

\begin{acks}
We are grateful to the anonymous reviewers for their valuable guidance and insightful comments. This paper is supported by the National Natural Science Foundation of China (Grant No. 62502467). 
\end{acks}
\bibliographystyle{ACM-Reference-Format}
\bibliography{references}

\newpage
\appendix
\section{Prompts for Global Context Summarization and Hierarchical Summary Generation} \label{code_summary}
\begin{figure}[b]
    \centering
    \includegraphics[width=\linewidth]{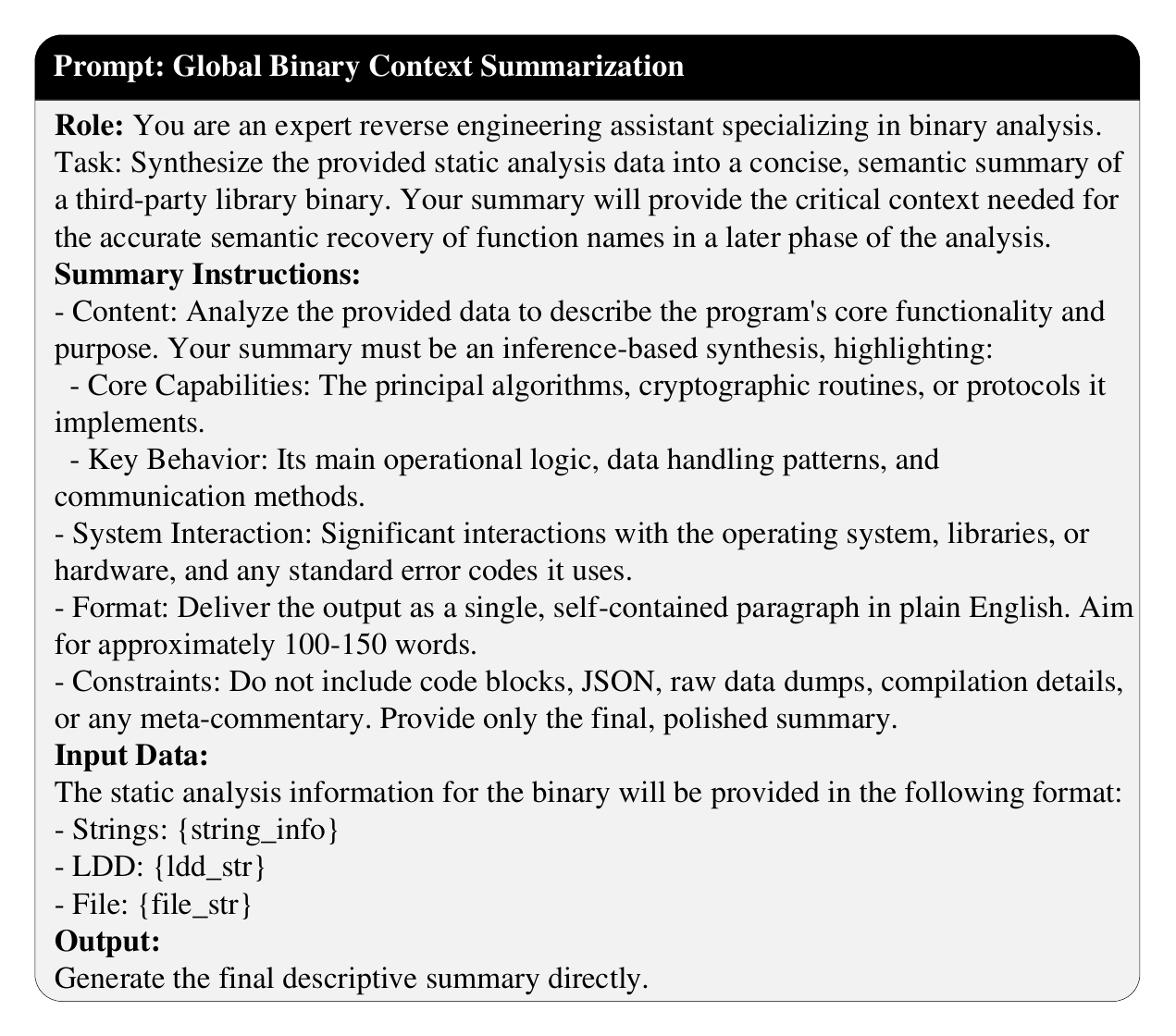}
    \caption{Prompt Template for Global Binary Context Summarization}
    \label{fig:prompt6}
    \vspace{-10pt}
\end{figure}

\begin{figure}[b]
    \centering
    \includegraphics[width=\linewidth]{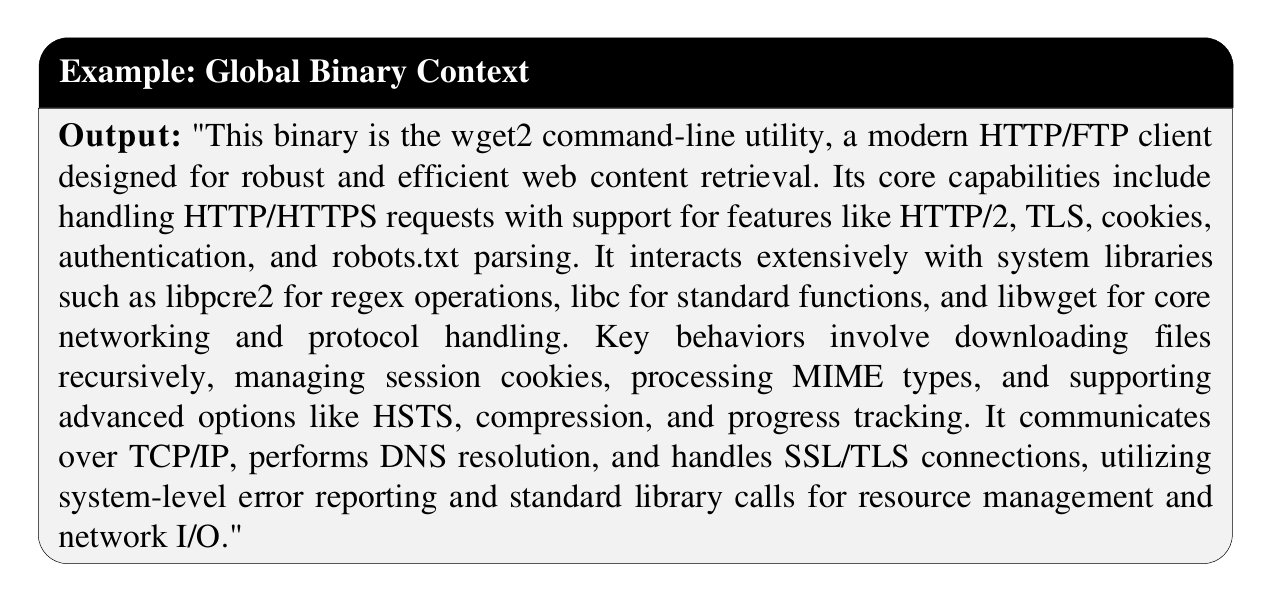}
    \caption{Example for Global Binary Context}
    \label{fig:Example Binary Information}
    \vspace{-6pt}
\end{figure}

Prompt design plays a critical role in determining the quality of LLM outputs. To generate global binary information as well as function-level and statement-level summaries, we experimented with multiple prompt variants on a small-scale validation dataset and ultimately selected the prompt configurations that yielded the best empirical performance.

For generating task‑related global binary context, we construct an input that includes a set of string literals extracted from the stripped binary, file-type metadata, and dynamic linkage dependencies. This information is provided to a LLM, which is instructed to produce a 100–150-word textual summary that captures the program’s core functionality and overall application context. The corresponding prompt template is shown in Figure~\ref{fig:prompt6}, and an example of the resulting global binary context is presented in Figure~\ref{fig:Example Binary Information}.

\begin{figure}[tb]
    \centering
     \vspace{-4pt}
    \includegraphics[width=\linewidth]{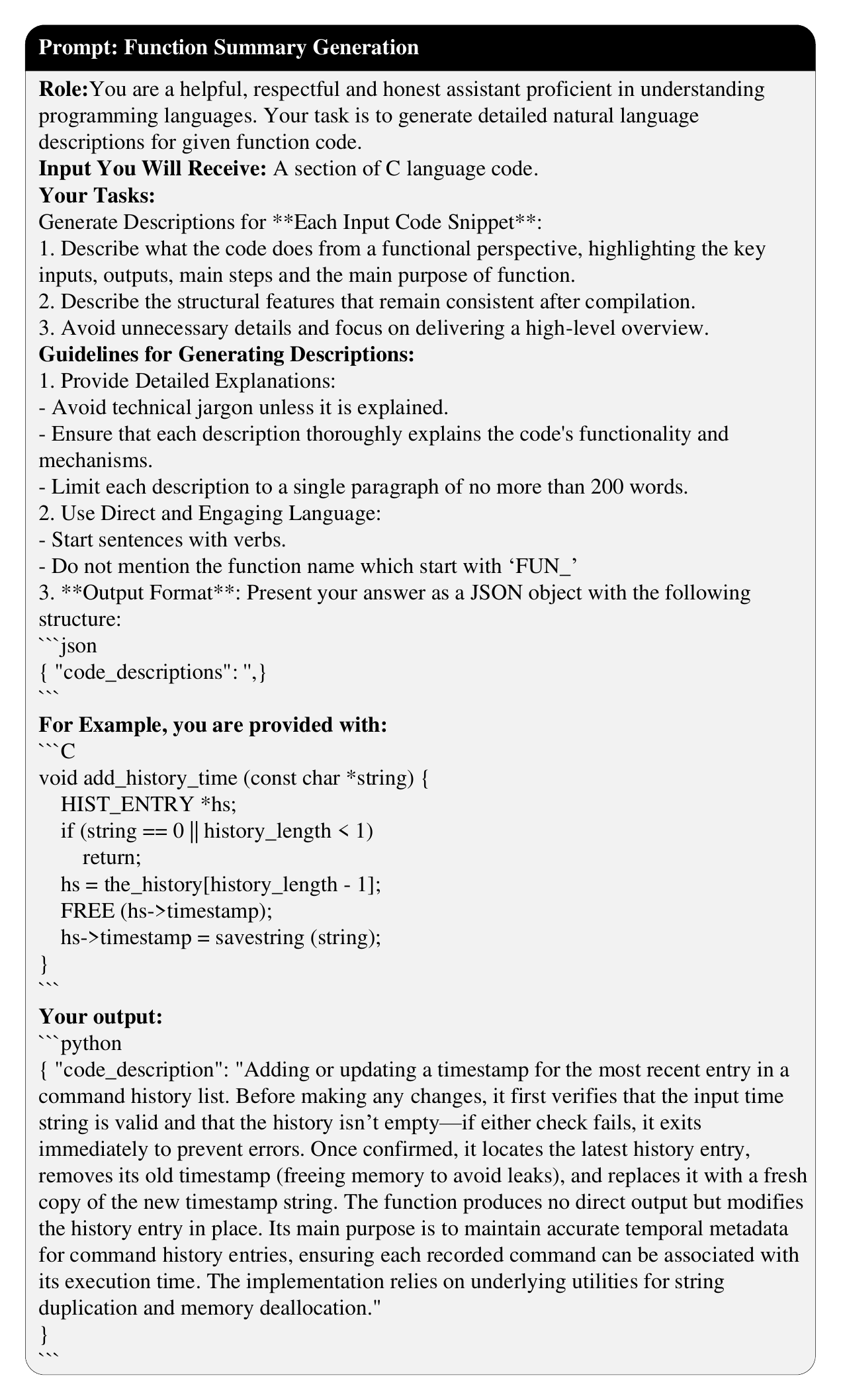}
    \caption{Prompt for Function Summary Generation}
    \label{fig:prompt1}
    \vspace{-4pt}
\end{figure}

\begin{figure}[tb]
    \centering
    \includegraphics[width=\linewidth]{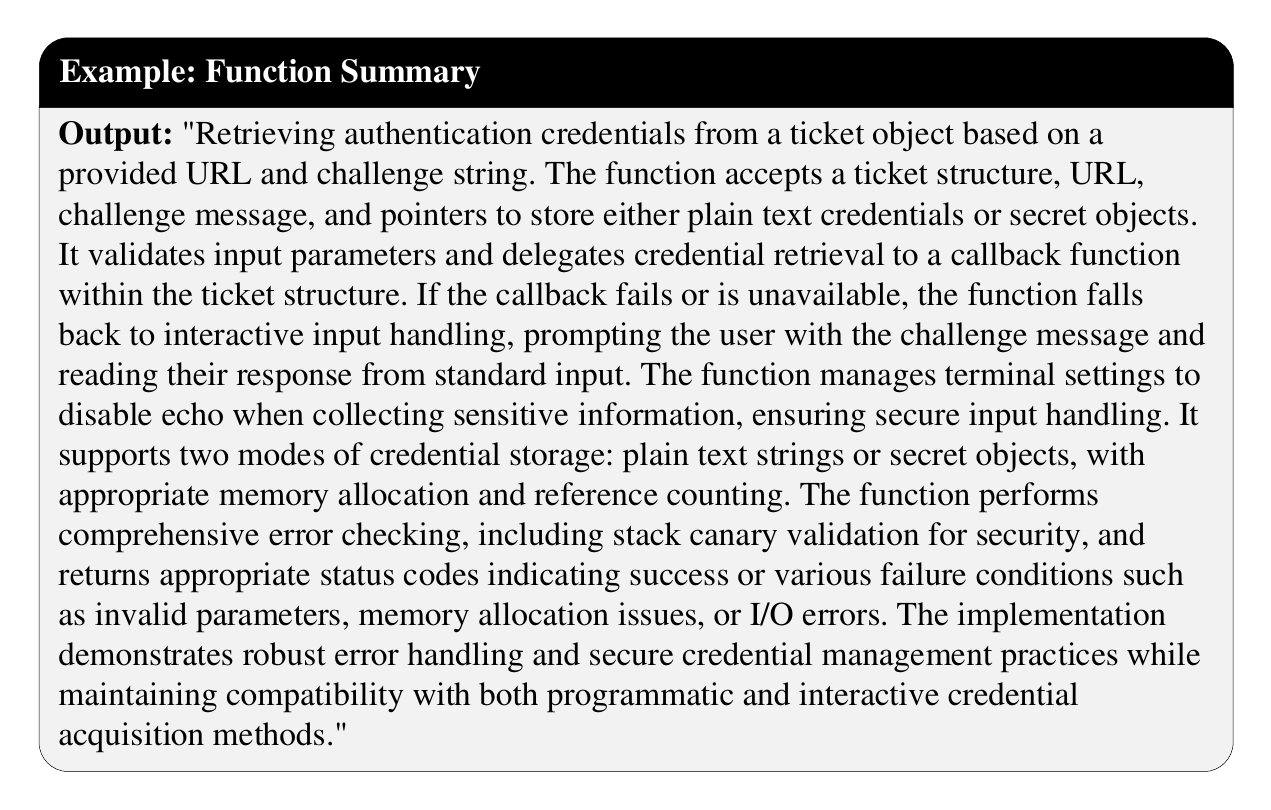}
    \caption{Example for Function Summary}
    \label{fig:Example Function Summary}
    \vspace{-10pt}
\end{figure}

\begin{figure}[tb]
    \centering
    \includegraphics[width=\linewidth]{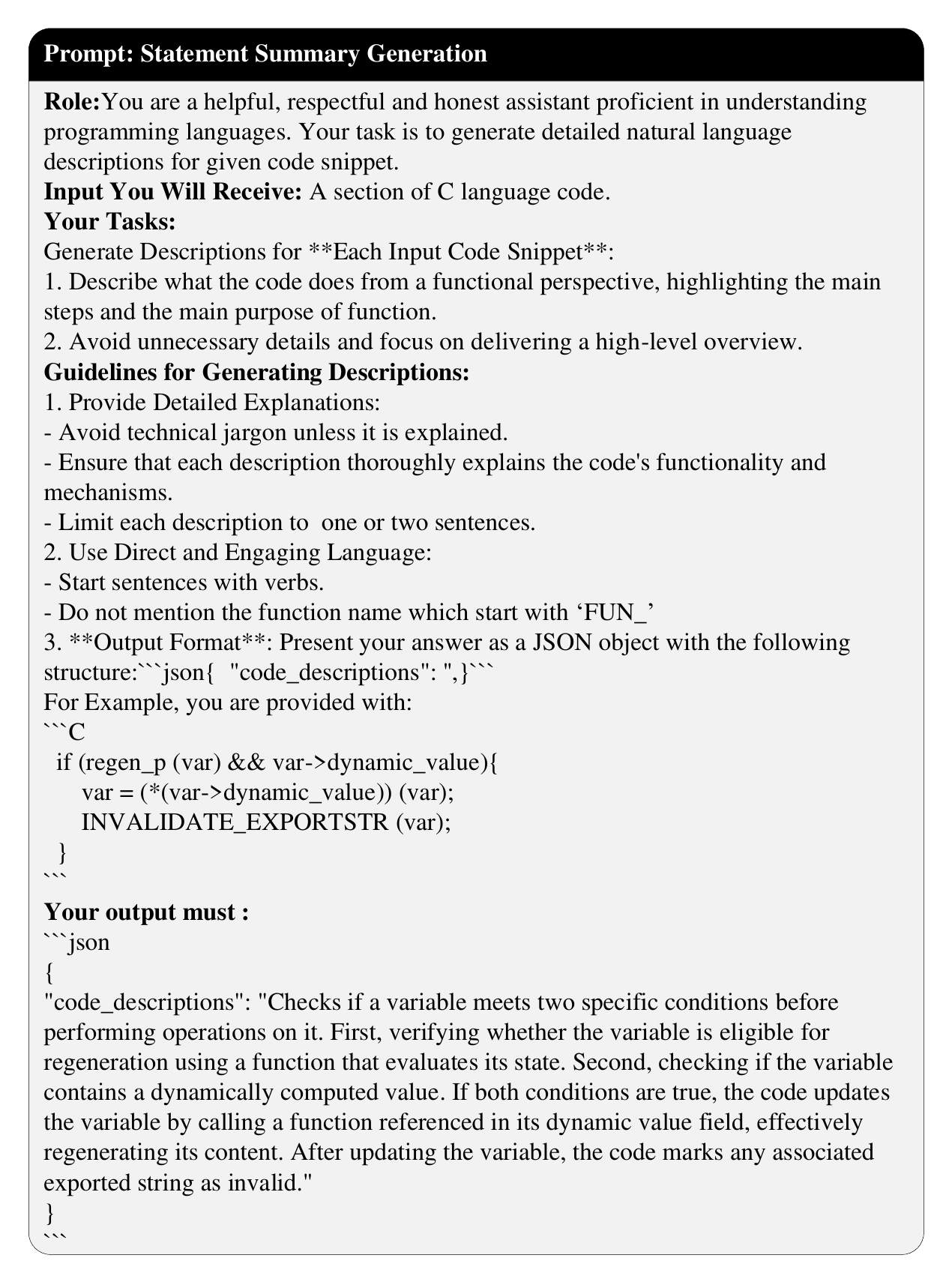}
    \caption{Prompt for Statement Summary Generation}
    \label{fig:prompt2}
\end{figure}

\begin{figure}[tb]
    \centering
    \vspace{-8pt}
    \includegraphics[width=\linewidth]{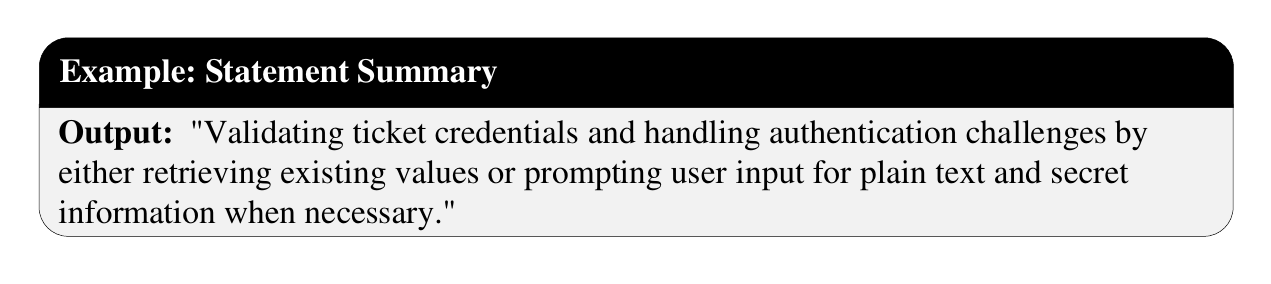}
    \caption{Example for Statement Summary}
    \label{fig:Example Statement Summary}
\end{figure}

The prompt used for function-level summary generation is illustrated in Figure~\ref{fig:prompt1}. In this setting, the original function name in the decompiled code is replaced with a special \texttt{[MASK]} token, while the function body, parameters, and other structural elements remain unchanged. This design encourages the model to infer function semantics solely from its implementation rather than relying on name cues. An example of a generated function-level summary is shown in Figure~\ref{fig:Example Function Summary}.
Similarly, the prompt for statement-level summarization, shown in Figure~\ref{fig:prompt2}, is specifically crafted to help \textsc{Hieronym} capture fine-grained semantic information at the statement level. An example of a statement-level summary is provided in Figure~\ref{fig:Example Statement Summary}. Together, these prompts enable the LLM to generate both high-level functional intent summary and low-level implementation logic summary.
To further guide the model toward accurately understanding the task objectives, both function-level and statement-level prompts adopt a one-shot format, in which a complete input–output example is provided. This in-context demonstration constrains the LLM’s generation behavior and steers it toward producing summaries that better match the desired format and semantic granularity.

\section{Prompt for Function Renaming Task Fine-Tuning} \label{fine_tuning_prompt}
\begin{figure}[tb]
    \centering
    \includegraphics[width=\linewidth]{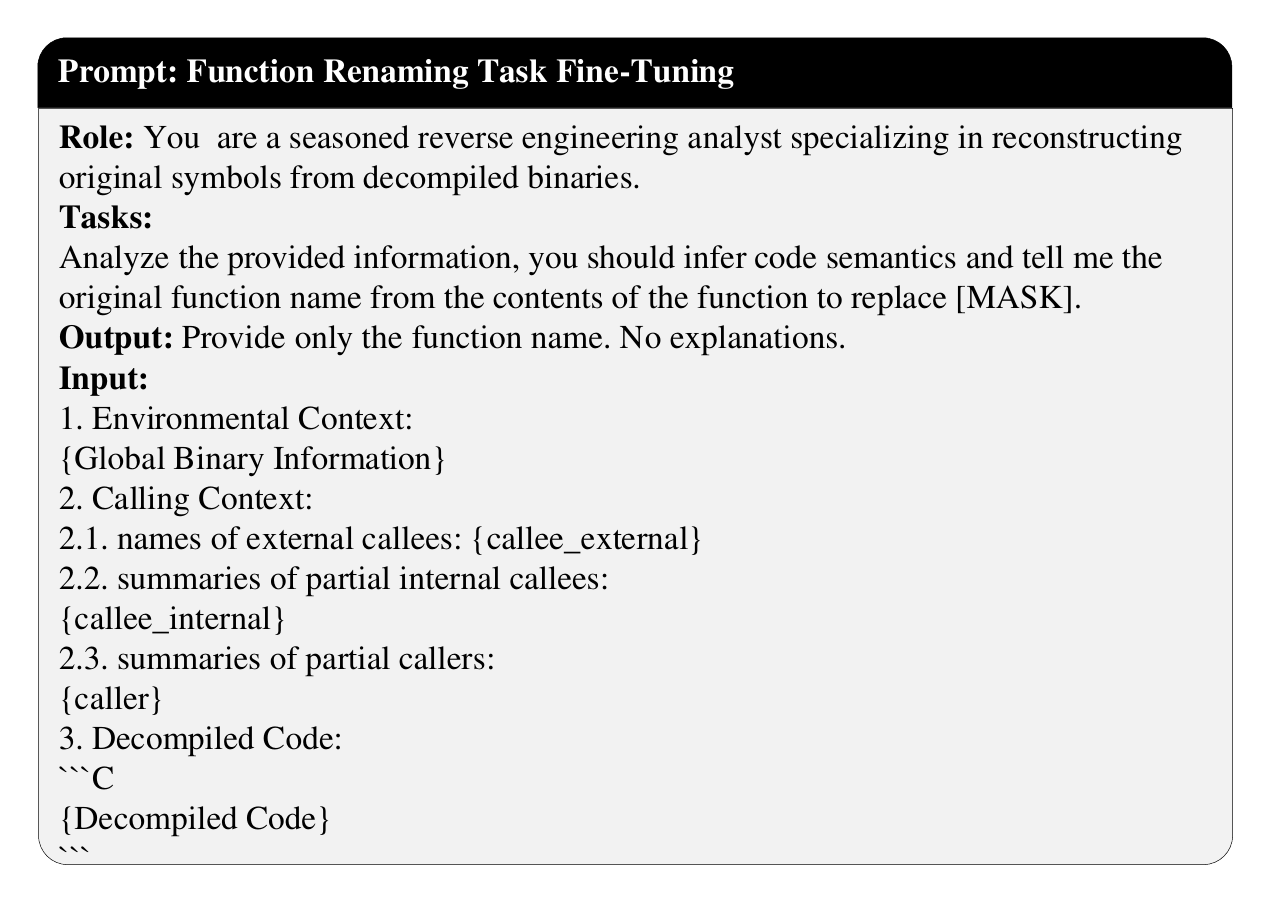}
    \caption{Prompt for Function Renaming Task Fine-Tuning}
    \label{fig:prompt4}
\end{figure}

For fine-tuning the LLM on the function renaming task for stripped binaries, we employ the prompt template shown in Figure~\ref{fig:prompt4}. The prompt input consists of three components: (i) global binary context; (ii) local calling context, where external callees are represented solely by their function names, while internal callers and callees are represented by their summaries (for efficiency, only the five most salient summaries are retained); and (iii) decompiled code of the target function, processed in the same manner as in function-level summarization, with the original function name replaced by a \texttt{[MASK]} token.
Consistent with our summarization tasks, we evaluated multiple prompt templates on a small-scale dataset and ultimately selected the configuration that achieved the best empirical performance for formal fine-tuning.

\section{Prompts for Evaluation} \label{tokenization}

To comprehensively evaluate the performance of function renaming methods, we adopt two complementary evaluation strategies: token-level evaluation and name–level evaluation.

\begin{figure}[H]
    \centering
    \includegraphics[width=\linewidth]{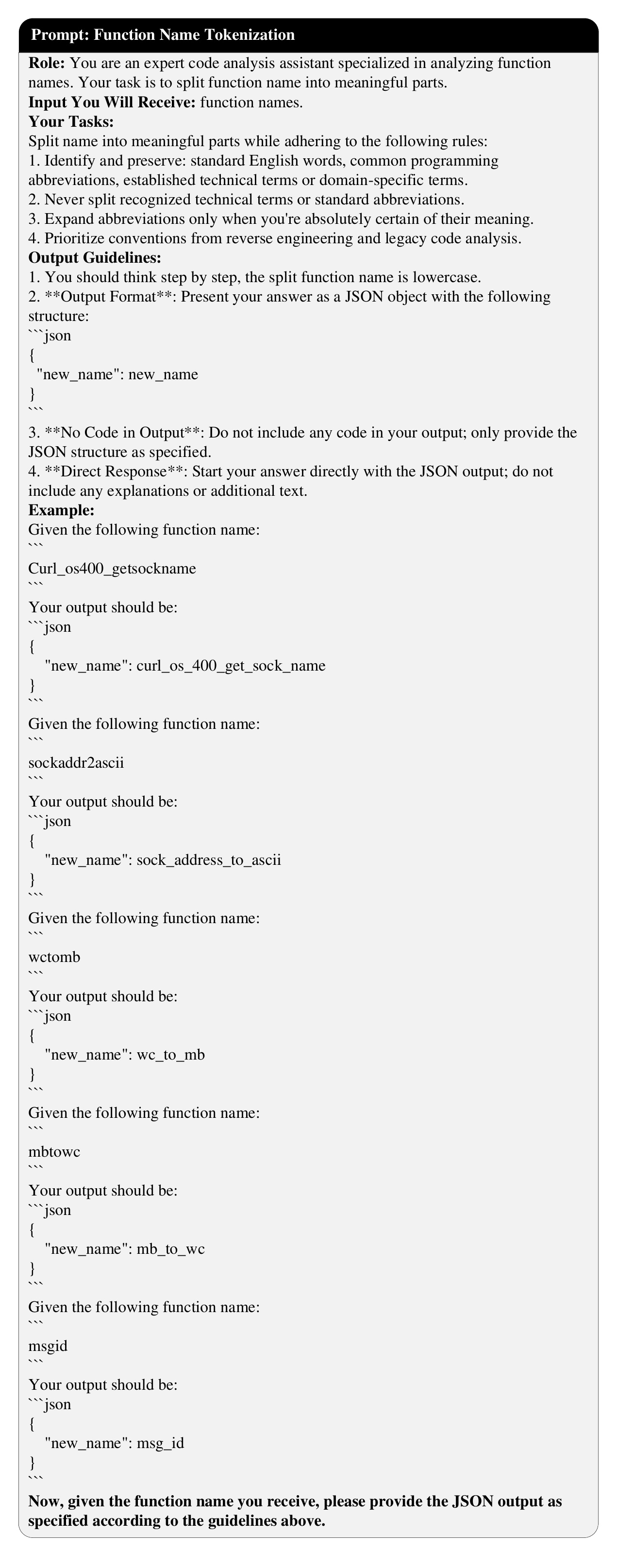}
    \caption{Prompt for Function Name Tokenization}
    \label{fig:prompt5}
\end{figure}

For token-level evaluation, function names are tokenized into semantically meaningful tokens. We leverage the contextual reasoning capabilities of LLMs and employ few-shot prompting to guide the tokenization process, the corresponding prompt design is illustrated in Figure~\ref{fig:prompt5}. This approach enforces segmentation only when semantic boundaries can be identified with high confidence, thereby reducing erroneous splits that would otherwise introduce noise into the evaluation. 

\begin{figure}[tb]
    \centering
    \includegraphics[width=\linewidth]{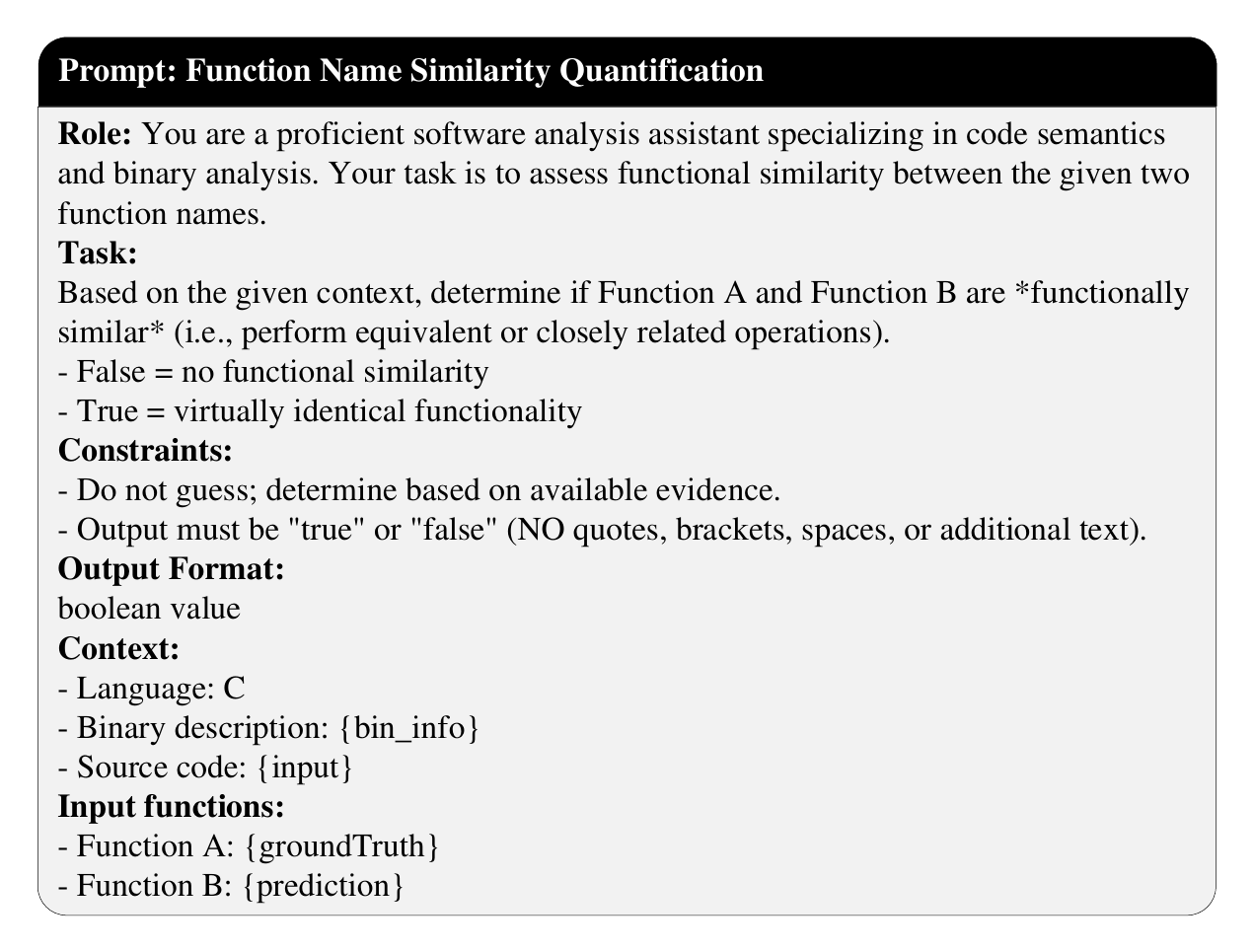}
    \caption{Prompt for Function Name Similarity Quantification}
    \label{fig:prompt8}
\end{figure}

For name–level evaluation, we exploit the discriminative capabilities of LLMs to determine whether a predicted function name is semantically consistent with the ground-truth name, given the binary’s global context and the decompiled code of the target function. The prompt template used for this evaluation is shown in Figure~\ref{fig:prompt8}. This strategy emphasizes holistic semantic alignment, is insensitive to incorrect tokenization, and more directly reflects the practical plausibility and usefulness of predicted function names in real-world analysis.

\begin{table*}[!t]
\centering
\caption{Source Projects and Binaries in Our Dataset}
\label{tab:dataset_projects}
\resizebox{1.7\columnwidth}{!}{
\begin{tabular}{l c c l c c}
\toprule
\textbf{Project} & \textbf{\# Binaries(GCC)} & \textbf{\# Binaries(Clang)} &\textbf{Project} & \textbf{\# Binaries(GCC)} & \textbf{\# Binaries(Clang)} \\
\midrule
    openssl-3.0.6        & 2128 &2164& libiconv-1.17        & 48 &40\\
    coreutils-8.32       & 1848 &1736& grep-3.8             & 48 &16\\
    ncurses-6.3          & 1440 &1168& gmp-6.2.1            & 40 &60\\
    readline-8.2         & 568  &440& libidn2-2.3.4        & 32 &28\\
    bash-5.2             & 472  &64& tar-1.34             & 32 &32\\
    adns-1.6.0           & 448  &256& datamash-1.8         & 32 &32\\
    gettext-0.21         & 444  &436& curl-7.86.0          & 32 &28\\
    inetutils-2.4        & 400  &380& gss-1.0.4            & 32 &28\\
    binutils-2.39        & 352  &452& poke-2.4             & 32 &88\\
    dico-2.11            & 276  &216& libpng-1.6.39        & 32 &148\\
    gawk-5.2.1           & 240  &216& gzip-1.12            & 16 &16\\
    libredwg-0.12        & 192  &168& libmicrohttpd-0.9.75 & 16 &16\\
    mailutils-3.8        & 180  &1288& libunistring-1.0     & 16 &16\\
    wget2-2.0.1          & 136  &44& cflow-1.7            & 16 &16\\
    freeipmi-1.6.10      & 112  &280& libtool-2.4.7        & 16 &16\\
    diffutils-3.8        & 96   &64& units-2.22           & 16 &16\\
    texinfo-7.0          & 54   &32&                      &    &\\
\bottomrule
\end{tabular}
}
\end{table*}

\section{Source Project and Binaries in Our Dataset} \label{symgen_dataset}

In our experiments, we adopt the SymGen dataset, which comprises 33 open-source projects and 9,842 unique binaries compiled with GCC 9.4.0 across four optimization levels (O0–O3) and four architectures (x86, x64, ARM, and MIPS). As shown in Table~\ref{tab:dataset_projects}, the dataset includes representative GNU projects such as bash, wget, and coreutils, which are widely used in binary analysis research.
To further evaluate the cross-compiler generalization of \textsc{Hieronym}, we recompile these projects using Clang 14.0.0 across the same optimization levels and architectures, producing an additional 10,000 unique binaries.

\section{Re-evaluation Based on SymGen Metrics} \label{symgen_eval}
To ensure fair comparison, we re-evaluate our results using the evaluation metrics adopted by SymGen. Specifically, SymGen splits both ground-truth and predicted function names into tokens and computes standard precision, recall, and F1-score metrics. Both exact matches and semantically similar tokens are treated as correct, which is consistent with our token-level evaluation framework.
To mitigate label noise caused by semantically related words, SymGen further employs CodeWordNet, introduced in SymLM~\cite{symlm}, which leverages word embeddings to capture relationships among morphologically or semantically related tokens. In contrast, our approach adopts a multi-dimensional association mechanism to identify semantically equivalent tokens, improving robustness to lexical variability.
Table~\ref{tab:overall_symgen} presents the experimental results across different architectures and optimization levels under SymGen’s evaluation setting. Even under this metric, \textsc{Hieronym} consistently outperforms SymGen, achieving improvements of 53.23\%, 44.33\%, and 47.96\% in precision, recall, and F1-score, respectively.
\begin{table}[tb]
    \centering
    \caption{Overall Performance of \textsc{Hieronym} and SymGen Across Architectures(Arch) and Optimizations(Opt)}
    \begin{center}
    \resizebox{\columnwidth}{!}{
    \begin{tabular}{cc|ccc|ccc}
    \toprule
         \multirow{2}{*}{\textbf{Arch}}&\multirow{2}{*}{\textbf{Opt}}&\multicolumn{3}{|c}{\textbf{\textsc{Hieronym}}}&\multicolumn{3}{|c}{\textbf{SymGen}}\\
          \cline{3-8}
         &&\textbf{Precision}&\textbf{Recall}&\textbf{F1-score}&\textbf{Precision}&\textbf{Recall}&\textbf{F1-score}\\
         \midrule
         \multirow{4}{*}{x64} &{O0}&0.4894 & 0.5262&0.5072&0.3547&0.3560&0.3553\\
         {}&{O1}&0.4926&0.4932&0.4929&0.3602 &0.3468&	0.3534\\
         {}&{O2}&0.4595&0.4745&0.4669&0.3683&	0.3360&	0.3514\\
         {}&{O3}&0.4668&0.4808&0.4737&0.3582&	0.3615&	0.3599 \\
         \hline
         \multirow{4}{*}{x86} &{O0}&0.4202& 	0.3435& 	0.3780 &0.3152& 	0.2806& 	0.2969\\
         {}&{O1}&0.4242& 	0.3940& 	0.4086& 0.3408& 	0.2926& 	0.3149 \\
         {}&{O2}&0.4103&0.3900&0.3999 &0.3167&0.2848&0.2999\\
         {}&{O3}&0.4269& 0.4177& 0.4223&0.3313&0.3135&0.3222 \\
         \hline
         \multirow{4}{*}{ARM} &{O0}&0.3364&0.3348 &0.3356&0.2799&0.2729&0.2764\\
         {}&{O1}&0.4003&0.3977&0.3990&0.2339&0.2129&0.2229\\
         {}&{O2}&0.3962& 	0.3937& 	0.3949&0.2795& 	0.2648&	0.2719\\
         {}&{O3}&0.3507& 	0.3483& 	0.3495&0.2448& 	0.2273& 	0.2357\\
         \hline
         \multirow{4}{*}{MIPS} &{O0}&0.3327& 	0.2501& 	0.2856 &0.1750& 	0.1893& 	0.1819 \\
         {}&{O1}&0.3376& 	0.2702& 	0.3002 &0.1628& 	0.1865& 	0.1739 \\
         {}&{O2}&0.3514& 	0.2692& 	0.3048&0.1713& 	0.1681& 	0.1697 \\
         {}&{O3}&0.2730& 	0.2364& 	0.2534 &0.1204& 	0.1397& 	0.1293 \\
    \bottomrule
    \end{tabular}
    }
    \label{tab:overall_symgen}
    \end{center}
\end{table}

\section{Cross-Compiler Generalization Results} \label{cross_com_result}
To evaluate the cross-compiler generalization of \textsc{Hieronym}, we re-evaluate both \textsc{Hieronym} and SymGen on binaries compiled with Clang. To ensure fair comparison, the experiments adopt the same test split and strict deduplication strategy as Section~\ref{data_setup} to prevent data leakage, and are evaluated using token-level metrics. 

Table~\ref{tab:overall_cross_compiler} summarizes the overall results across different architectures and optimization levels. \textsc{Hieronym} achieves a weighted macro-average precision of 0.4694, recall of 0.4499, and F1-score of 0.4587, while SymGen attains 0.3344 precision, 0.3028 recall, and 0.3173 F1-score, corresponding to relative improvements of 47.46\%, 52.19\%, and 49.68\%, respectively. 
On the original GCC-compiled test set, \textsc{Hieronym} achieves precision, recall, and F1-score of 0.4487, 0.4223, and 0.4339, respectively, compared to 0.3180, 0.3018, and 0.3093 for SymGen. The comparable performance across compilers indicates strong cross-compiler generalization for both methods, likely because LLM-based approaches operate on decompiled code, which reduces compiler-specific variation. Notably, \textsc{Hieronym} further benefits from its model design and consistently achieves superior naming performance.
In addition, both methods exhibit architecture-dependent trends on the dataset compiled with Clang similar to those observed on the GCC-compiled dataset: performance remains strong on x64 and x86, but drops substantially on ARM and MIPS. 

\begin{table}[tb]
    \centering
    \caption{Cross-Compiler Performance of \textsc{Hieronym} and SymGen Across Architectures(Arch) and Optimizations(Opt)}
    \begin{center}
    \resizebox{\columnwidth}{!}{
        \begin{tabular}{cc|ccc|ccc}
        \toprule
         \multirow{2}{*}{\textbf{Arch}}&\multirow{2}{*}{\textbf{Opt}}&\multicolumn{3}{|c}{\textbf{\textsc{Hieronym}}}&\multicolumn{3}{|c}{\textbf{SymGen}}\\
          \cline{3-8}
         &&\textbf{Precision}&\textbf{Recall}&\textbf{F1-score}&\textbf{Precision}&\textbf{Recall}&\textbf{F1-score}\\
         \midrule
         \multirow{4}{*}{x64}
        {}&{O0} & 0.5947 & 0.6362 & 0.6148 & 0.4044 & 0.3172 & 0.3556 \\
        {}&{O1} & 0.5813 & 0.6087 & 0.5947 & 0.4803 & 0.4599 & 0.4699 \\
        {}&{O2} & 0.5788 & 0.6204 & 0.5989 & 0.4621 & 0.4098 & 0.4344 \\
        {}&{O3} & 0.5817 & 0.6308 & 0.6053 & 0.4318 & 0.3930 & 0.4115 \\
        \hline
        \multirow{4}{*}{x86}
        {}&{O0} & 0.5286 & 0.4726 & 0.4990 & 0.4377 & 0.3793 & 0.4064 \\
        {}&{O1} & 0.5616 & 0.5078 & 0.5333 & 0.4207 & 0.3614 & 0.3888 \\
        & O2 & 0.5524 & 0.4984 & 0.5240 & 0.4192 & 0.3703 & 0.3932 \\
        & O3 & 0.5423 & 0.4879 & 0.5137 & 0.4236 & 0.4009 & 0.4119 \\
        \hline
        \multirow{4}{*}{ARM}
        {}&{O0} & 0.3876 & 0.3463 & 0.3658 & 0.2947 & 0.2792 & 0.2867 \\
        {}&{O1} & 0.3890 & 0.3457 & 0.3661 & 0.2318 & 0.1894 & 0.2085 \\
        & O2 & 0.3820 & 0.3462 & 0.3632 & 0.2903 & 0.2496 & 0.2684 \\
        & O3 & 0.3805 & 0.3386 & 0.3583 & 0.2665 & 0.2202 & 0.2412 \\
        \hline
        \multirow{4}{*}{MIPS}
        {}&{O0}& 0.3692 & 0.3390 & 0.3534 & 0.2156 & 0.2269 & 0.2211 \\
        {}&{O1} & 0.3601 & 0.3441 & 0.3519 & 0.1881 & 0.1922 & 0.1901 \\
        & O2 & 0.3584 & 0.3367 & 0.3472 & 0.2107 & 0.2028 & 0.2067 \\
        & O3 & 0.3629 & 0.3388 & 0.3505 & 0.1722 & 0.1931 & 0.1821 \\
        \bottomrule
        \end{tabular}
    }
    \label{tab:overall_cross_compiler}
    \end{center}
\end{table}

\section{Cross-Decompiler Generalization Results} \label{cross_decom_result}
To evaluate the cross-decompiler generalization of \textsc{Hieronym}, we redecompile the test set using IDA Pro 9.0 and re-evaluate both \textsc{Hieronym} and SymGen on the resulting dataset. The experiments also adopt the same strict deduplication strategy as Section~\ref{data_setup} to prevent data leakage, and performance is evaluated using token-level metrics.

Table~\ref{tab:overall_cross_decompiler} summarizes the results across architectures and optimization levels. Overall, \textsc{Hieronym} achieves a weighted macro-average precision of 0.4867, recall of 0.4433, and F1 of 0.4633, while SymGen attains 0.4219 precision, 0.3557 recall, and 0.3853 F1, corresponding to relative improvements of 15.56\%, 25.21\%, and 20.54\%, respectively. 
For comparison, on the original GCC-compiled test set, \textsc{Hieronym} achieves precision, recall, and F1 scores of 0.4487, 0.4223, and 0.4339, respectively, whereas SymGen achieves 0.3180, 0.3018, and 0.3093.
These results indicate strong cross-decompiler generalization for both methods, with SymGen exhibiting particularly large gains. Further analysis shows that the improvement mainly arises from ARM and MIPS binaries. Manual inspection reveals that IDA Pro can often recover referenced string information directly within decompiled ARM functions, whereas Ghidra typically represents the same content as memory addresses. Such recovered strings provide valuable semantic cues for function name prediction. 
Because \textsc{Hieronym} already incorporates global and call-context information in its design, which supplies additional semantic cues, the contribution of recovered strings is comparatively limited, although improvements on ARM and MIPS are still observed. In contrast, SymGen lacks such multi-source contextual inputs and therefore benefits more substantially from the additional string information. These findings suggest that improving decompilation quality is an effective direction for enhancing function renaming performance.

\begin{table}[tb]
    \centering
    \newcommand{\uparrowgreen}{\textcolor[rgb]{0,0.85,0}{\uparrow}}
    \caption{Cross-Decompiler Performance of \textsc{Hieronym} and SymGen Across Architectures(Arch) and Optimizations(Opt)}
    \begin{center}
    \resizebox{\columnwidth}{!}{
    \begin{tabular}{cc|ccc|ccc}
    \toprule
         \multirow{2}{*}{\textbf{Arch}}&\multirow{2}{*}{\textbf{Opt}}&\multicolumn{3}{|c}{\textbf{\textsc{Hieronym}}}&\multicolumn{3}{|c}{\textbf{SymGen}}\\
          \cline{3-8}
         {}&{}&{\textbf{Precision}}&{\textbf{Recall}}&{\textbf{F1-score}}&{\textbf{Precision}}&{\textbf{Recall}}&{\textbf{F1-score}}\\
         \midrule
        \multirow{4}{*}{x64}
        & O0 & 0.4758 & 0.4772 & 0.4765 & 0.4262 & 0.3514 & 0.3852 \\
        & O1 & 0.4808 & 0.4781 & 0.4794 & 0.4269 & 0.3388 & 0.3778 \\
        & O2 & 0.4899 & 0.4764 & 0.4830 & 0.4195 & 0.3280 & 0.3681 \\
        & O3 & 0.4803 & 0.4855 & 0.4829 & 0.4228 & 0.3392 & 0.3764 \\
        \hline
        \multirow{4}{*}{x86}
        & O0 & 0.4867 & 0.4165 & 0.4489 & 0.4379 & 0.3716 & 0.4020 \\
        & O1 & 0.4753 & 0.3793 & 0.4219 & 0.4500 & 0.3419 & 0.3886 \\
        & O2 & 0.4851 & 0.3899 & 0.4323 & 0.4301 & 0.3108 & 0.3608 \\
        & O3 & 0.4926 & 0.4120 & 0.4487 & 0.4625 & 0.3861 & 0.4209 \\
        \hline
        \multirow{4}{*}{ARM}
        & O0 & 0.5059 & 0.4332 & 0.4668 & 0.4419 & 0.3928 & 0.4159 \\
        & O1 & 0.5112 & 0.4495 & 0.4784 & 0.3961 & 0.3190 & 0.3534 \\
        & O2 & 0.5021 & 0.4439 & 0.4712 & 0.4149 & 0.3566 & 0.3835 \\
        & O3 & 0.5005 & 0.4440 & 0.4706 & 0.4141 & 0.3607 & 0.3856 \\
        \hline
        \multirow{4}{*}{MIPS}
        & O0 & 0.4997 & 0.4673 & 0.4830 & 0.4164 & 0.3954 & 0.4056 \\
        & O1 & 0.4814 & 0.4586 & 0.4698 & 0.4056 & 0.3725 & 0.3858 \\
        & O2 & 0.4779 & 0.4481 & 0.4625 & 0.3984 & 0.3556 & 0.3758 \\
        & O3 & 0.4423 & 0.4333 & 0.4378 & 0.3875 & 0.3711 & 0.3791 \\
    \bottomrule
    \end{tabular}
    }
    \label{tab:overall_cross_decompiler}
    \end{center}
\end{table}
\end{document}